\documentclass[reprint,aps,prd,amssymb, amsmath, nofootinbib]{revtex4-2}

\usepackage{graphicx}

\usepackage{mathtools}
\usepackage[svgnames]{xcolor} 

\usepackage[colorlinks=True, linkcolor=SlateBlue,
            citecolor=SteelBlue,urlcolor=BlueViolet]{hyperref}
\usepackage[capitalise]{cleveref}
\usepackage{orcidlink}

\usepackage{aas_macros}

\newcommand{\vect}[1]{\boldsymbol{#1}}
\newcommand{\dvect}[1]{\dot{\boldsymbol{#1}}}
\newcommand{\uvect}[1]{\vect{\hat{#1}}}
\newcommand{\duvect}[1]{\vect{\dot{\hat{#1}}}}

\newcommand{\uLN}{\uvect{L}_{\rm N}}
\newcommand{\chieff}{\chi_{\rm eff}}

\newcommand{\xphm}{\texttt{IMRPhenomXPHM}}
\newcommand{\xode}{\texttt{IMRPhenomXODE}}
\newcommand{\seob}{\texttt{SEOBNRv4PHM}}
\newcommand{\LAL}{\texttt{LAL}}
\newcommand{\LALS}{\texttt{LALSuite}}

\newcommand{\teja}[2]{\textcolor{black}{#1}}

\defcitealias{Pratten:21}{G. Pratten \textit{et al.}, Phys. Rev. D 103, 104056 (2021)}
\defcitealias{Ossokine:20}{S. Ossokine \textit{et al.}, Phys. Rev. D 102, 044055 (2020)}

\begin{document}

\author{Hang Yu\,\orcidlink{0000-0002-6011-6190}}
\email{hang.yu2@montana.edu}

\affiliation{Kavli Institute for Theoretical Physics, University of California at Santa Barbara, Santa Barbara, CA 93106, USA}
\affiliation{eXtreme Gravity Institute, Department of Physics, Montana State University,
Bozeman, Montana 59717, USA}

\author{Javier Roulet}
\affiliation{TAPIR, Walter Burke Institute for Theoretical Physics, Mailcode 350-17\\ California Institute of Technology, Pasadena, CA 91125, USA}

\author{Tejaswi Venumadhav}
\affiliation{Department of Physics, University of California at Santa Barbara, Santa Barbara, CA 93106, USA}
\affiliation{International Centre for Theoretical Sciences, Tata Institute of Fundamental Research, Bangalore 560089, India}

\author{Barak Zackay}
\affiliation{Department of Particle Physics and Astrophysics, Weizmann Institute of Science, Rehovot 76100, Israel}

\author{Matias Zaldarriaga}
\affiliation{School of Natural Sciences, Institute for Advanced Study, Princeton, New Jersey 08540, USA}


\title{Accurate and Efficient Waveform Model for Precessing Binary Black Holes}

\begin{abstract}
We present \xode, a new phenomenological frequency-domain waveform approximant for gravitational wave (GW) signals from precessing binary black holes (BBHs) with generic spin configurations. We build upon the success of \xphm~[\citetalias{Pratten:21}], which is one of the most widely adopted waveform approximants in GW data analyses that include spin precession, and introduce two additional significant improvements. First, we employ an efficient technique to numerically solve the (next-to)$^4$-leading-order post-Newtonian precession equations, which allows us to accurately determine the evolution of the orientation of the orbital angular momentum $\uLN$ even in cases with complicated precession dynamics, such as transitional precession. Second, we recalibrate the phase of GW modes in the frame coprecessing with $\uLN$ against \seob~[\citetalias{Ossokine:20}] to capture effects due to precession such as variations in the spin components aligned with $\uLN$. By incorporating these new features, \xode\ achieves matches with \seob\ that are better than 99\% for most BBHs with mass ratios $q \geq 1/6$ and with arbitrary spin configurations. In contrast, the mismatch between \xphm\ and \seob\ often exceeds 10\% for a BBH with $q\lesssim 1/2$ and large in-plane or antialigned spin components. Our implementation is also computationally efficient, with waveform evaluation times that can even be shorter than those of \xphm\ for BBH signals with long durations and hence high frequency resolutions. The accuracy and efficiency of \xode\ position it as a valuable tool for GW event searches, parameter estimation analyses, and the inference of underlying population properties.

\end{abstract}

\maketitle

\section{Introduction}

To date, gravitational-wave (GW) detectors including Advanced LIGO (aLIGO;~\cite{LSC:15det}) and Advanced Virgo~\cite{Acernese:15} have enabled the detection of $\sim 100$ mergers of binary black holes (BBHs) by both the LIGO-Virgo-KAGRA (LVK) collaboration~\cite{LVC:19gwtc1, LVC:21gwtc2, LVC:21gwtc2.1, LVK:21gwtc3} and other independent groups~\cite{Venumadhav:20, Olsen:22, Nitz:19, Nitz:20, Nitz:21, Nitz:23}. 
Detailed parameter estimation and population studies show that many of the detected BBH signals are consistent with the component black holes having no significant spin angular momentum, but there are several events that show signs of at least one of the merging black holes having non-zero spins \cite{Zackay:19, Biscoveanu:21, Roulet:21, Tong:22, MateuLucena:22, LVK:23pop}. 
This motivates the development of waveform models that can describe signals from BBH systems that host spinning components. 

The dominant effect of spins is to modify the phase evolution of the gravitational waveforms in the inspiral region \cite{Kidder:93}. Right from the early days of gravitational wave searches for compact binary coalescences (CBC), this effect has been included in waveform models \cite{Cutler:94, Poisson:95, Buonanno:03, Ajith:11}. Spins can also modify the dynamics of the subsequent merger phase, and detailed Inspiral-Merger-Ringdown (IMR) waveforms include corrections to account for this \cite{Campanelli:06, Ajith:11b}. 

Gravitational wave signals from CBCs can show an especially rich phenomenology if the components have spins that are misaligned with the orbital angular momentum. In this case, the BBH's orbital plane will experience general relativistic precession and nutation, leaving detectable modulations in the phase and amplitude of the GW signal~\cite{Apostolatos:94, Gangardt:21}. Some potentially precessing systems detected during LVK's third observing run include GW190412~\cite{LVC:20GW190412}, GW190521~\cite{LVC:20GW190521}, GW191109\_010717~\cite{LVK:21gwtc3} and GW200129\_065458~\cite{Hannam:22}. Knowing the spin configuration of each individual event and the population properties of the spin distribution~\cite{LVK:21pop, LVK:23pop, Roulet:21, Tong:22, Perigois:23} is of great scientific interest because the spin encodes key information about the formation channels of the BBHs~\cite{Rodriguez:16, Stevenson:17, Talbot:17, Zaldarriaga:18, Kimball:20, Gompertz:22, Antonini:18, Liu:18, Yu:20a}, as well as angular momentum transport in the progenitor stars~\cite{deMink:09, Mandel:16, Fuller:19} and supernova physics~\cite{Chan:20}. 



Matched-filtering-based searches as well as parameter estimation routines work with the likelihood of the data given a set of signal parameters, which necessitates accurate and efficient waveform templates~\cite{Jaranowski:12, Cutler:94}. Waveform approximants for BBHs with spins aligned (or anti-aligned) with the orbital angular momentum have reached great sophistication and excellent accuracy (see, e.g., Refs.~\cite{Buonanno:09, Husa:16, Khan:16, Cotesta:18, Pratten:20, GarciaQuiros:20, Cotesta:20}), but approximants for precessing BBHs still have room for improvement, with major developments having happened relatively recently over the past decade. 
Broadly speaking, the existing waveform approximants for precessing BBHs are constructed based on three types of methods: surrogate models of numerical relativity (NR)~\cite{Field:14, Blackman:15, Blackman:17, Varma:19}, effective-one-body (EOB) theory~\cite{Ossokine:20, Akcay:21, Gamba:22}, and Phenomenological models~\cite{Schmidt:12, Hannam:14, Khan:19, Khan:20, Pratten:21, Estelles:22, Hamilton:21}. 

Our most accurate description of strong-field gravitational interaction comes from NR simulations~\cite{Boyle:19, Hamilton:23b}. Surrogate waveform models directly approximate the outputs of these simulations in a data-driven manner, without appealing to post-Newtonian theory except for generating the overall intuition for the approach. NR simulations typically cover only the last few cycles of the inspiral followed by the final merger-ringdown. Surrogate waveforms are trained based on the NR results~\cite{Blackman:15}, and hence only have a small frequency coverage and cannot be directly used for analyzing the data from arbitrarily low frequencies to capture the early inspiral phase. A promising method to address this is to smoothly transition from EOB waveforms to surrogate waveforms as the system evolves from the early inspiral to the merger-ringdown part, a process which is termed {\em hybridization}. This has been successfully implemented for non-precessing mergers~\cite{Varma:19b}.

The effective-one-body approach aims to bridge the post-Newtonian (PN) regime and the NR regime by adding parameters to the description of the merging system that are tuned, or {\em calibrated}, to reproduce the behavior from simulations~\cite{Ossokine:20, Akcay:21, Gamba:22}. This formalism involves both our best analytical knowledge for the inspiral-plunge part and phenomenological calibrations to NR in the merger-ringdown part.  
Some of the latest developments in the EOB family for precessing BBHs include \seob~\cite{Ossokine:20} and \texttt{TEOBResumS}~\cite{Akcay:21, Gamba:22}. These models are constructed in the time domain and have demonstrated good accuracy when tested against NR simulations. Each evaluation of an effective-one-body waveform requires the solution of a set of ordinary differential equations that describe the orbital evolution of the binary, as well as the component spins. Additionally, analyses typically transform the data and the signal into the frequency domain, which adds an extra step on top. In order to mitigate these computational costs, reduced order models have been developed to directly approximate the EOB waveforms using singular value decomposition \cite{Purrer:14, Purrer:16}, reduced order quadrature \cite{Smith:16}, and other machine-learning-based approaches \cite{Schidt:21, Thomas:22}.

Phenomenological waveform models~\cite{Hannam:14, Khan:19, Khan:20, Pratten:21,  Hamilton:21} are among the most widely used models in practical data analyses as they are typically constructed directly in the frequency domain (except for Ref.~\cite{Estelles:22}) and are computationally efficient to evaluate. Phenomenological waveforms can be evaluated at arbitrary sets of frequencies without first solving for their form on a finer regular frequency grid, which makes them especially useful for heterodyning or relative-binning \cite{Cornish:10, Cornish:21, Zackay:18}, which is a general method to speed up parameter inference.

\xphm~is a state-of-the-art frequency-domain phenomenological waveform model that includes the effect of orbital precession in CBCs with misaligned spins, as well as higher harmonics~\cite{Pratten:21}. As with many other approximants, \xphm\ constructs the waveform using a ``twisting-up" technique~\cite{Schmidt:12}. The approach is to first model the waveform in a frame that precesses along with the orbital angular momentum vector (the so-called coprecessing frame), and then apply a time-dependent frame rotation to convert the waveform to the inertial frame of a distant observer. In \xphm, the coprecessing waveform is modeled as a non-precessing one based on \texttt{IMRPhenomXAS}~\cite{Pratten:20}  and \texttt{IMRPhenomXHM}~\cite{GarciaQuiros:20} with an update of the final spin. More realistically, the frequency evolution of the waveform even in the coprecessing frame should be coupled with the precession dynamics (as implemented by, e.g., \seob~\cite{Ossokine:20} at the expense of a much heavier computational cost). Furthermore, the precession dynamics are modeled either using a single-spin PN expansion (as used in \texttt{IMRPhenomPv2}~\cite{Hannam:14}) or using a multi-scale analysis~\cite{Chatziioannou:17} that assumes a hierarchy of timescales between orbital precession and the orbital evolution (similar to the one adopted by \texttt{IMRPhenomPv3HM}~\cite{Khan:20}). Both approximants can give good, but not exact representations of the precession.\footnote{During the preparation of this manuscript, a new upgrade of \xphm, named \xphm-\texttt{SpinTaylor}~\cite{Colleoni:23} has been implemented in \LALS~\cite{LAL:18} and is newly available for parameter estimation during the fourth observing run of LVK. The new implementation utilizes the exact numerical solutions to the \texttt{SpinTaylor} infrastructure~\cite{Sturani:19} 
to improve the accuracy of the precession dynamics compared to the analytical approximations.} 
For example, neither approximation can describe BBHs experiencing transitional precession~\cite{Apostolatos:94} and nutational resonances~\cite{Zhao:17}. 

We introduce \xode, a new frequency domain approximant that addresses the above issues. Our approximant is based on \xphm\ but extends it in two aspects. First of all, we track the orientation of the orbital angular momentum by numerically solving the ordinary differential equations (ODEs) governing precession dynamics at the (next-to)$^4$ leading order (N4LO)~\cite{Akcay:21}. To maintain or even enhance the computational efficiency of the waveform generation, we use a change of variables that takes care of the leading evolution and helps accelerate the numerical integration of the differential equations for the precession dynamics. Second, we recalibrate the phase of the GW modes in the coprecessing frame against \seob~\cite{Ossokine:20} to capture variations in the phase evolution that are associated with the in-plane components of the spins.

Throughout this paper, we will use geometrical units with $G=c=1$. When referring to a BBH, we denote the more massive component as $M_1$ and the other $M_2$ so that the mass ratio $q=M_2/M_1\leq1$. The total mass will be denoted by $M=M_1+M_2$. Other useful quantities we will use frequently include the symmetric mass ratio $\eta=M_1 M_2/M^2$ and the orbital angular frequency $\omega$. The velocity is thus $v=(M\omega)^{1/3}$. Unless otherwise stated, we will refer to $f=\omega/\pi$ as the GW frequency, though one should keep in mind that different modes [labeled by angular quantum numbers $(\ell,\ m)$], in general, will hit the same GW frequency (defined based on the instantaneous time derivative of the mode's phase) at different values of $\omega$, and we have $f=\omega/\pi$ only for the $|m|=2$ modes including the dominant $\ell=2$ quadrupole mode. 
When going from the time domain to the frequency domain, we adopt the following Fourier transform convention, 
\begin{equation}
    \tilde{h}(f) = \int h(t) e^{-2\pi i ft} dt. 
\end{equation}
This is consistent with the convention adopted by the LIGO Algorithm Library, \LALS~\cite{LAL:18, SWIGLAL:20}. 

The rest of the paper is organized as follows. In Sec.~\ref{sec:overview}, we introduce the reference frame adopted by our study and present an overview of the waveform construction procedure using the twisting-up technique. In Sec.~\ref{sec:precession} we describe the precession dynamics and the associated impacts on the GW waveform. Specifically, the precession ODEs we solve are presented in Sec.~\ref{sec:prec_EoM}. The limitation of the multiple-scale analysis is discussed in Sec.~\ref{sec:limitations_p_avg}, which is followed by Sec.~\ref{sec:fast_prec} where we describe the techniques we employ to accelerate the evaluation of precession ODEs. The recalibration of coprecessing GW waveforms is in Sec.~\ref{sec:cp_modes}. After describing the ingredients entering our new waveform, we systematically compare the matches between \xode, \xphm, and \seob\ in Sec.~\ref{sec:validation}. We also comment on the reasons that make it impractical to use NR to validate the approximant. Lastly, we conclude and discuss the results in Sec.~\ref{sec:conclusion_discussions}. 

\section{Reference frames and overview of the waveform construction}
\label{sec:overview}

We follow the literature~\cite{Boyle:11, Schmidt:12, Hannam:14, Khan:20, Ossokine:20, Akcay:21, Pratten:21} and build the \xode\ approximant using the ``twisting-up'' construction: \teja{we first construct the waveform in the coprecessing frame that tracks the orbital angular momentum, and then rotate to an inertial frame using Euler angles that track the precession dynamics}{the waveform is first constructed in a frame coprecessing with the orbital angular momentum and then rotated to an inertial frame in terms of Euler angles that track the precession dynamics}. 

The reference frames in \xode\ follow the default convention adopted by \xphm~\cite{Pratten:21}. To describe the problem, we refer to the coprecessing frame as the $L$-frame, since its $z$-axis tracks the instantaneous orientation of the Newtonian orbital angular momentum $\uLN$. We also introduce a $J$-frame whose $z$-axis is aligned with the instantaneous total angular momentum $\vect{J} = \vect{L} + \vect{S}_1 + \vect{S}_2$, where $\vect{S}_{1(2)}$ is the spin of $M_{1(2)}$ and $\vect{L}$ is related to its Newtonian value through~\cite{Bohe:13, Akcay:21} 
\begin{align}
    \vect{L} &= \frac{\eta}{v} \left\{
        \uLN \left[
            1 + v^2 \left(\frac{3}{2} + \frac{\eta}{6}\right)
              + v^4 \left(\frac{27}{8}-\frac{19\eta}{8} + \frac{\eta^2}{24}\right)
        \right]\nonumber 
    \right.\\
    &\qquad~
    \left.
        {}+ v^3 \Delta \vect{L}_{\rm 1.5PN}^{S} + v^5 \Delta \vect{L}_{\rm 2.5PN}^{S}
    \right\},
\end{align}
where $\Delta \vect{L}_{n {\rm PN}}^{S}$ were spin corrections for which we take the orbital-averaged value from Ref.~\cite{Akcay:21} [see their eqs. (A4) and (A5)]. 
Note that while the precession-averaged orientation of $\vect{J}$ is approximately a constant, precession averaging \teja{can be inaccurate}{may not be accurate} for many BBHs in the LIGO--Virgo sensitivity band (see Sec.~\ref{sec:limitations_p_avg})\teja{ and hence, we cannot treat the $J$ frame as an inertial frame in general.}{. Therefore, the $J$ frame is not an inertial frame in general.} Nonetheless, we can introduce an inertial $J_0$ frame that coincides with the $J$ frame at a reference GW frequency $f_{\rm ref} \equiv \omega_{\rm ref}/\pi$. Similarly, an inertial $L_0$ frame is introduced to coincide with the $L$ frame at $f_{\rm ref}$. 
Following the \LAL\ convention, the line-of-sight vector $\uvect{N}$ is given in the $L_0$ frame as~\cite{Schmidt:16, Pratten:21}
\begin{equation}
    \uvect{N} = 
    \begin{pmatrix}
    \sin \iota \,\cos(\pi/2-\phi_{\rm ref}) \\
    \sin \iota \,\sin(\pi/2-\phi_{\rm ref}) \\
    \cos \iota
    \end{pmatrix}_{L_0},
\end{equation}
where $\iota$ and $\phi_{\rm ref}$ are input parameters specified by the user as in \xphm. 
Components of the spins at $f_{\rm ref}$ are also input from the user in the $L_0$ frame.
In particular, we will use $(\chi_{1x}, \chi_{1y}, \chi_{1z})$ to denote the components of dimensionless spin $\vect{\chi}_1=\vect{S}_1/M_1^2$ of BH $M_1$ in the $L_0$ frame at $f_{\rm ref}$, and similarly for the spin of $M_2$. We will further define $\chi_{1p}=\sqrt{\chi_{1x}^2 + \chi_{1y}^2}$ to characterize the spin component perpendicular to the Newtonian orbital angular momentum. 

The azimuthal orientation of the $J_0$ frame is fixed by requesting $\uvect{N}$ to be in the $xz$-plane, so that
\begin{equation}
    \uvect{N} = 
    \begin{pmatrix}
    \sin \theta_{JN}, \\
    0, \\
    \cos \theta_{JN}
    \end{pmatrix}_{J_0}.
    \label{eq:alphadef}
\end{equation}
To rotate the coprecessing waveform to the inertial frame, Euler angles $(\alpha, \beta)$ are defined through the components of $\uLN$ in the $J_0$ frame, 
\begin{equation}
    \uLN = 
    \begin{pmatrix}
    \sin \beta \,\cos \alpha, \\
    \sin \beta \,\sin \alpha, \\
    \cos \beta
    \end{pmatrix}_{J_0}.
    \label{eq:def_euler_angs}
\end{equation}
The last Euler angle $\epsilon$ is defined through $\dot{\epsilon} = \dot{\alpha} \cos \beta $. The waveform in the $J_0$ frame is now given in terms of the one in the coprecessing frame, and the Euler angles $(\alpha, \beta, \epsilon)$ as~\cite{Schmidt:11} 
\begin{equation}
    h_{\ell m}^{J_0}=e^{-im\alpha}\sum_{m'}e^{im'\epsilon} d_{mm'}^{\ell}(\beta) h_{\ell m'}^L,
    \label{eq:h_J_lm_from_h_L_lm}
\end{equation}
where we have decomposed the waveform into spin-weighted spherical harmonics ${}_{-2} Y_{\ell m}$ specified with quantum numbers $(\ell, m)$ and $d_{mm'}^\ell$ denote the real-valued Wigner-$d$ matrices.
Note that we can completely absorb a constant offset in $\epsilon$ by a constant phase shift of $h_{\ell m'}^L$. In our code, we adopt the default convention of the \xphm\ code; see appendix C of Ref.~\cite{Pratten:21} for details. 

The two polarizations of the GW are related to the modes via~\cite{Arun:09}
\begin{equation}
    h_+^{J_0} - i h_\times^{J_0} = \sum_{\ell, m} h^{J_0}_{\ell m} {}_{-2} Y_{\ell m}(\theta_{JN}, \phi_{JN}). 
\end{equation}
Note $\phi_{JN} = 0$ in our convention, as $\uvect{N}$ is in the $xz$-plane in the $J_0$ frame.  
We can further rotate the polarization to a frame set by $\uvect{N}$. The details are described in appendices C and D of Ref.~\cite{Pratten:21}. 

We performed the decomposition in Eq.~\eqref{eq:h_J_lm_from_h_L_lm} in the time domain. The Euler angles change on a precession timescale (see Sec.~\ref{sec:limitations_p_avg}) that is much longer than the gravitational-wave period, and hence we can adopt a stationary phase approximation~\cite{Cutler:94} to write the frequency-domain waveform as~\cite{Pratten:21} 
\begin{align}
    &\tilde{h}_+^{J_0}(f>0)=\frac{1}{2}\sum_\ell \sum_{m'<0} e^{im' \epsilon} \tilde{h}_{\ell m}^L(f)\nonumber \\
        \times & \sum_m \left[
            e^{-i m\alpha} d_{mm'}^\ell(\beta) {}_{-2}Y_{\ell m} + 
            e^{i m \alpha} d_{m,-m'}^\ell(\beta) {}_{-2}Y_{\ell m}^\ast
        \right], 
    \label{eq:hlm_L_2_hp_J}
    \\
    &\tilde{h}_\times^{J_0}(f>0)=\frac{i}{2}\sum_\ell \sum_{m'<0} e^{im' \epsilon} \tilde{h}_{\ell m}^L(f)\nonumber \\
        \times & \sum_m \left[
            e^{-i m\alpha} d_{mm'}^\ell(\beta) {}_{-2}Y_{\ell m} - 
            e^{i m \alpha} d_{m,-m'}^\ell(\beta) {}_{-2}Y_{\ell m}^\ast
        \right],
    \label{eq:hlm_L_2_hc_J}
\end{align}
In the expressions above, we evaluate $\alpha=\alpha[\omega(f)]$ and similarly for $\beta$ and $\epsilon$. In other words, we treat the Euler angles as functions of the orbital frequency and then express the orbital frequency in terms of the GW frequency $f$. For a specific coprecessing mode with an azimuthal quantum number $m'$, the two are approximately related by $\omega \simeq 2\pi f/|m'|$ (note only $m'<0$ modes have support for $f>0$ in the \LAL~convention). The correction due to the GW tail~\cite{Arun:09, Boyle:11, Blanchet:14} is included in the code, but its effect is negligible in practice. 

We now see that the construction of the final waveform requires two ingredients, the evolution of $(\alpha, \beta, \epsilon)$ and the coprecessing modes $\tilde{h}_{\ell m}^{L}$. We will describe how we obtain each one respectively in Secs.~\ref{sec:precession} and \ref{sec:cp_modes}.

\section{Precession dynamics}
\label{sec:precession}

From Eq.~(\ref{eq:def_euler_angs}) we see that the evolution of the Euler angles is determined once we obtain the evolution of $\uLN$, which is the main focus of this Section. In particular, we present the defining equations governing the precession dynamics in Sec.~\ref{sec:prec_EoM}. 
\teja{In the literature, approximate analytical solutions to the set of equations have been derived using a multi-scale analysis (MSA; \cite{Chatziioannou:17}) based on precession averaging. We discuss the limitations of this approximation in Sec.~\ref{sec:limitations_p_avg}.}{While approximate analytical solutions to the set of equations have been derived using a multi-scale analysis (MSA; \cite{Chatziioannou:17}) based on precession averaging, we discuss in Sec.~\ref{sec:limitations_p_avg} the limitations of such approximation.} Lastly, we introduce a technique that allows us to obtain the \emph{exact} solutions of the precession equations with high efficiency. 

\subsection{Equations of motion}
\label{sec:prec_EoM}

In this work, we follow the equations governing the precession dynamics given in Ref.~\cite{Akcay:21}. 
At the next-to-leading order (NLO), we have~\cite{Kesden:15, Chatziioannou:17}
\begin{align}
    \duvect{S}_1^{\rm (NLO)} &= v^5 \eta \left(2+\frac{3}{2} q\right)\left(\uLN \times \uvect{S}_1\right) \nonumber \\
    &\quad+ \frac{v^6}{2} \left\{\vect{S}_2 - 3 \left[(q\vect{S}_1 + \vect{S}_2)\cdot \uLN \right]\uLN\right\} \times \uvect{S}_1,  
    \label{eq:dS1dt_NLO}\\
    \duvect{S}_2^{\rm (NLO)} &= v^5 \eta \left(2+\frac{3}{2q} \right)\left(\uLN \times \uvect{S}_2\right) \nonumber \\
    &\quad+ \frac{v^6}{2} \left\{\vect{S}_1 - 3 \left[(\vect{S}_1 + \vect{S}_2/q)\cdot \uLN \right]\uLN\right\} \times \uvect{S}_2, 
    \label{eq:dS2dt_NLO}\\
    \duvect{L}_{\rm N}^{\rm (NLO)} &= -\frac{v}{\eta}(\dvect{S}_1 + \dvect{S}_2).
    \label{eq:dLdt_NLO}
\end{align}
The full N4LO dynamics we adopt can be found in Ref.~\cite{Akcay:21}.\footnote{Note that while the N4LO equations include spin-orbit interactions at higher PN orders than the NLO equations do, the spin-spin interactions remain at the same PN order in both sets of equations. The impact due to neglecting higher-order spin-spin interactions in the N4LO dynamics remains to be addressed by future studies. }
Note that while we have $\duvect{L}_{\rm N}^{\rm (NLO)} \perp \uLN$, at N4LO $\duvect{L}_{\rm N}$ has components both perpendicular and parallel to $\uLN$. Consistent with Refs.~\cite{Sturani:19, Akcay:21}, we track only the perpendicular component of $\duvect{L}_{\rm N}$ in our code. 
GW decay enters implicitly via $v=(M\omega)^{1/3}$, whose evolution we model as~\cite{Chatziioannou:13} 
\begin{equation}
    \dot{\omega} = \frac{a_0 }{M^2} v^{11}
    \left[ 1+
    \sum_{i=2}^{7}\left(a_i+3 b_i\ln v\right) v^i
    \right]. 
    \label{eq:dwdt}
\end{equation}
The coefficients of $(a_i, b_i)$ are provided in appendix A of Ref.~\cite{Chatziioannou:13}. The values of $(\chi_{1z}, \chi_{2z})$ entering $(a_i, b_i)$ are updated based on the instantaneous $\uvect{S}_1 \cdot \uLN$ and $\uvect{S}_2 \cdot \uLN$ in the code by default. 
Since our waveform is constructed in the frequency domain, it is more convenient to express the evolution in terms of the orbital frequency $\omega$ instead of time $t$. We can accomplish this by dividing both sides of Eqs.~(\ref{eq:dS1dt_NLO})--(\ref{eq:dLdt_NLO}) by $\dot{\omega}$ given in Eq.~(\ref{eq:dwdt}). 

Strictly speaking, the equations are valid only in the inspiral regime. We nonetheless evolve them to a binary separation of $3M$ or a point when $\dot{\omega}=0$, and then fix the orientations of all the vectors afterward. \teja{}{} \teja{More sophisticated treatments of the Euler angles through the merger-ringdown parts have been proposed in the literature~\cite{Ossokine:20, Hamilton:21}, and for completeness, we should incorporate these effects. Our  treatment effectively freezes the evolution of the Euler angles through the merger, with the intuition that the dominant timescales at this point are much faster than that of precession. The results of Ref.~\cite{Gamba:22}, and our numerical comparisons in Sec.~\ref{sec:validation} show that the resulting loss of accuracy is under control.}{As we will see shortly in Sec.~\ref{sec:limitations_p_avg}, our treatment effectively freezes the evolution of the Euler angles after the merger. While more sophisticated treatments of the Euler angles in the merger-ringdown parts have been proposed~\cite{Ossokine:20, Hamilton:21}, Ref.~\cite{Gamba:22} suggests that simply fixing their values after the merger would give comparable accuracy.} 

Approximate analytical solutions to Eqs.~(\ref{eq:dS1dt_NLO})--(\ref{eq:dLdt_NLO}) have been derived. If only one BH is spinning, Eqs.~(\ref{eq:dS1dt_NLO})--(\ref{eq:dLdt_NLO}) can be readily solved under a PN expansion~\cite{Schmidt:12, Hannam:14}. When both BHs are spinning, one can map it to the single-spin case via introducing $\chi_p$, defined as~\cite{Schmidt:15} 
\begin{equation}
    \chi_p = \frac{1}{A_1 M_1^2}{\rm max}\left(A_1 M_1^2\chi_{1p}, A_2 M_2^2 \chi_{2p} \right), 
\end{equation}
where $A_1 = (2 + 3q/2)$, $A_2=(2+3/2q)$, and $\chi_{1p (2p)}$ is the magnitude of the component of $\chi_{1(2)}$ that is perpendicular to $\uLN$. The expansion at the next-next-to-leading order (NNLO) is adopted by \texttt{IMRPhenomPv2}~\cite{Hannam:14} and can be requested in \xphm.

A slightly more accurate description taking into account two-spin effects can be derived using an MSA~\cite{Chatziioannou:17} under precession averaging~\cite{Kesden:15}. This is the default prescription of precession in \xphm. However, as we will show in Sec.~\ref{sec:limitations_p_avg}, the MSA has its limitations and many of its underlying assumptions in fact break down for BBHs in the LIGO band.

\subsection{Limitations of the MSA construction}
\label{sec:limitations_p_avg}

\begin{figure}
  \centering
  \includegraphics[width=\columnwidth]{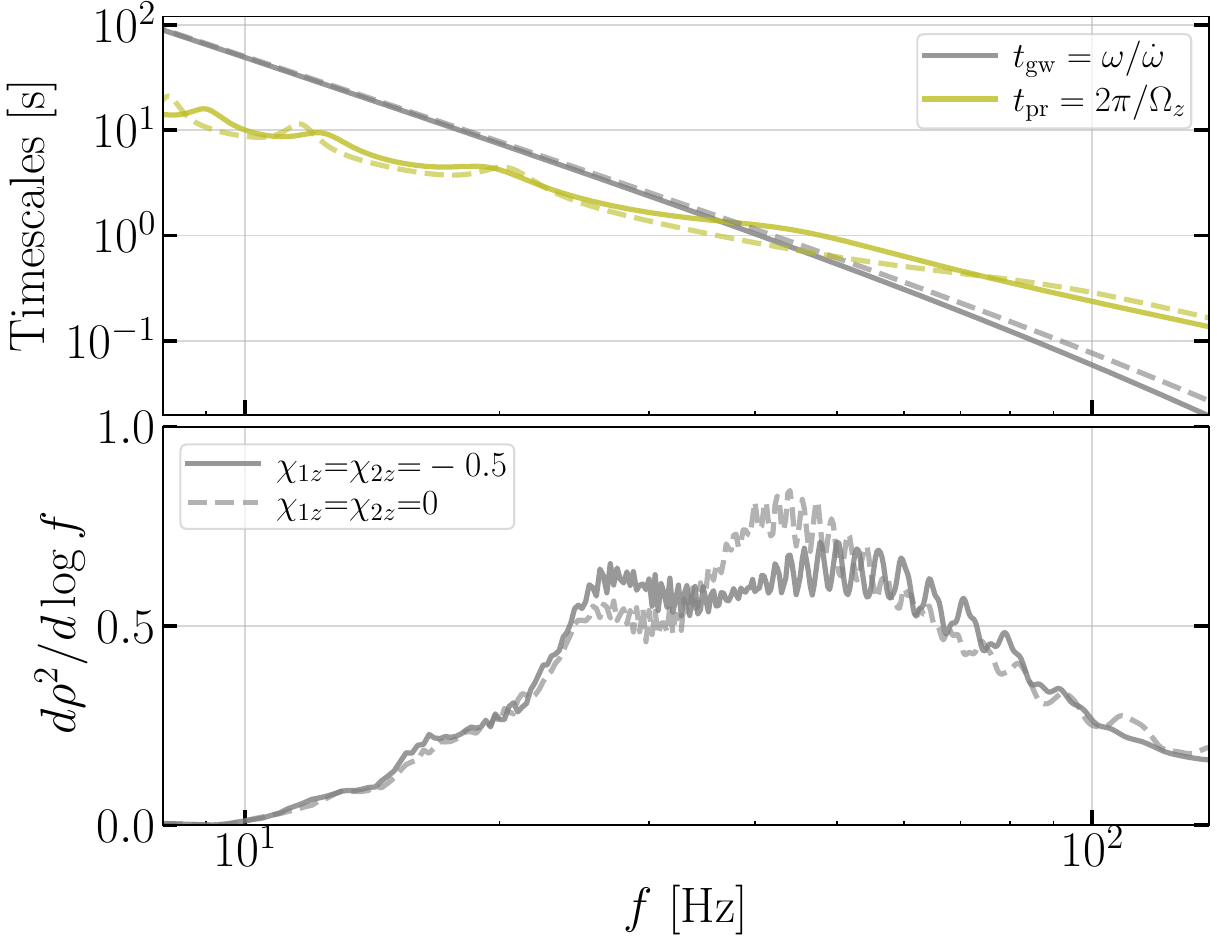}
\caption{{\em Top:} comparison between the precession timescale and the GW decay timescale. The system has $(M_1, M_2) = (20\,M_\odot, 12\,M_\odot)$ with two different spin configurations, $(\chi_{1z}, \chi_{1p}) = (\chi_{2z}, \chi_{2p})=(-0.5, 0.5)$ (solid lines) or $(\chi_{1z}, \chi_{1p}) = (\chi_{2z}, \chi_{2p})=(0, 0.5)$ (dashed lines). {\em Bottom:} Distribution of the squared signal-to-noise ratio $\rho^2$ per logarithmic frequency, with the overall normalization set to $\rho=1$. Note $t_{\rm pr}<t_{\rm gw}$ holds only for $f\lesssim 40\,{\rm Hz}$, which contributes only to about half of $\rho^2$.
}
\label{fig:timescale_comp}
\end{figure}

One obvious improvement over the MSA formulation is to incorporate precession dynamics at higher PN orders. The MSA formulation is derived for precession dynamics at NLO whereas N4LO equations are available and are incorporated in our waveform construction. 

Even restricting the precession dynamics to NLO, the MSA formulation still has a few limitations. 
One of the key assumptions behind the MSA construction is that there exists a separation of timescales, with the precession timescale much shorter than the GW-induced orbital decay timescale. This assumption enables the use of closed-form expressions derived in Ref.~\cite{Kesden:15} to efficiently track the precession dynamics at NLO. More quantitatively, the precession rate of the system (at NLO and ignoring GW decay) is~\cite{Kesden:15, Chatziioannou:17, Yu:20a}
\begin{align}
    \Omega_z &= \frac{J}{2}v^6\bigg\{ 1+\frac{3}{2\eta} (1-\chieff v) \nonumber \\
    &-\frac{3(1+q)}{2q A_1^2 A_2^2} (1-\chieff v)\Big[ 4(1-q)L^2 \left(S_1^2 - S_2^2\right) \nonumber \\
    &-(1+q)\left(J^2-L^2-S^2\right)\left(J^2-L^2-S^2-4\eta L \chieff\right) \Big] \bigg\}, 
\end{align}
where $S=|\vect{S}_1 + \vect{S}_2|$, $A_1^2 = J^2 - (L-S)^2$, $A_2^2 = (L+S)^2 - J^2$, and $\chi_{\rm eff}=(M_1\chi_{1z} + M_2 \chi_{2z})/M$. 
One can thus define a precession timescale 
\begin{equation}
    t_{\rm pr} = \frac{2\pi}{\Omega_z}.
\end{equation}
It is to be compared with the orbital decay timescale
\begin{equation}
    t_{\rm gw} = \frac{\omega}{\dot{\omega}}. 
\end{equation}
The assumed separation of timescales is based on the observation of $t_{\rm pr}\propto v^{-5}$ (assuming $J\sim L\simeq \eta M^2/v$) whereas $t_{\rm gw}\propto v^{-8}$.

However, such a comparison ignores the large numerical coefficients of the timescales. A slightly more careful comparison leads to $t_{\rm pr}/t_{\rm gw} \sim [384\pi/5(4+3q)] M\omega$ where we have simplified $\Omega_z$ according to Ref.~\cite{Apostolatos:94}. Therefore, at a binary separation of $6M$, typically we have $t_{\rm pr}/t_{\rm gw} \sim 2$--$4$ using only the leading-order terms. In other words, precession is in fact slower than orbital decay near the end of the inspiral (\teja{see also the discussion in Refs.~\cite{JohnsonMcDaniel:22} and \cite{Gerosa:23}}{}). 

We demonstrate the reversal of the timescale hierarchy further in Fig.~\ref{fig:timescale_comp}. 
Here we consider a BBH system with $(M_1, M_2) = (20\,M_\odot, 12\,M_\odot)$ with  two different spin configurations. The solid lines correspond to $(\chi_{1z}, \chi_{1p}) {=} (\chi_{2z}, \chi_{2p}){=}(-0.5, 0.5)$ and the dashed lines correspond to $(\chi_{1z}, \chi_{1p}) {=} (\chi_{2z}, \chi_{2p}){=}(0, 0.5)$. In both cases, the spins are specified at a reference frequency $f_{\rm ref}=4\,{\rm Hz}$.  
We terminate the plot when the binary hits a separation of $6M$.  In the top panel, we show $t_{\rm gw}$ and $t_{\rm pr}$ respectively in the gray and olive lines. We then present the density of the squared signal-to-noise ratio (SNR) $\rho^2$ of the system assuming aLIGO sensitivity~\cite{Barsotti:18} in the bottom panel (\teja{the squared SNR is an integral over frequency, so its density is a meaningful quantity to think about}{}; note that when precession is included, the amplitude of the waveform is in general not a smooth function due to the interference between modes in the inertial frame, causing wiggles in the SNR density). We have normalized the signal so that $\rho^2=1$. For such a BBH with parameters typical of what ground-based GW observatories are expected to detect, we see the assumption of $t_{\rm pr} < t_{\rm gw}$ breaks down at around $f\simeq 40-50\,{\rm Hz}$, where the SNR density of the system peaks. In fact, near the end of the inspiral stage, the precession timescale can be longer than the GW decay timescale by nearly an order of magnitude.

\begin{figure}
  \centering
  \includegraphics[width=\columnwidth]{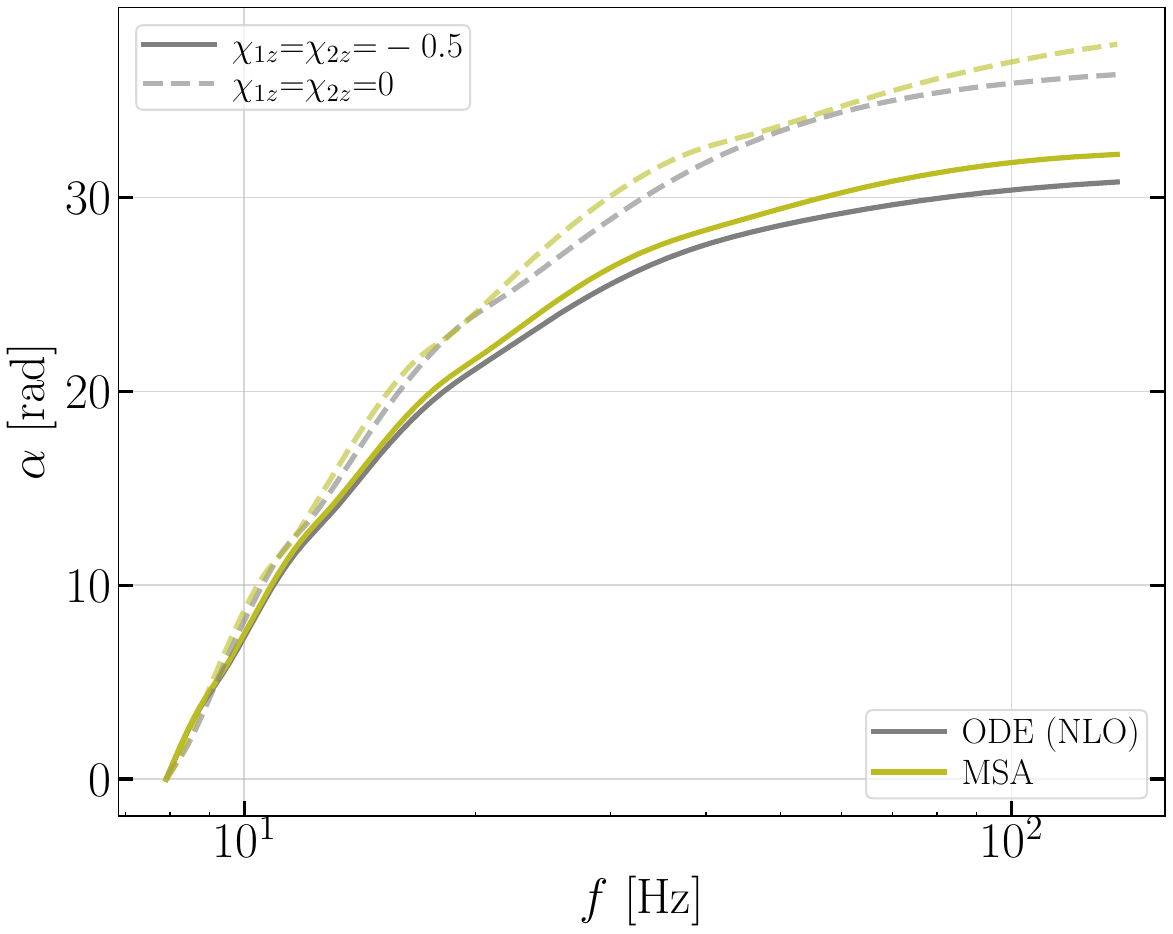}
\caption{Evolution of the Euler angle $\alpha$, defined in Eq.~\eqref{eq:def_euler_angs}, for the system shown in Fig.~\ref{fig:timescale_comp}. The MSA computation can lead to an error of $\sim 1\,{\rm rad}$ in $\alpha$ compared to the numerical solution to the ODE (at NLO to be consistent with the MSA derivation).}
\label{fig:MSA_vs_NLO_ODE}
\end{figure}

\teja{The Euler angles computed from the MSA can have non-negligible errors compared to the numerical solutions due to the breakdown of the assumption of a hierarchy in timescales. This is demonstrated in Fig.~\ref{fig:MSA_vs_NLO_ODE} (see also the top panel of Fig.~\ref{fig:XODE_vs_SEOB_debug})}{Because the assumption about timescale hierarchy breaks down, the Euler angles computed from the MSA can have non-negligible errors compared to the numerical solutions which we demonstrate in Fig.~\ref{fig:MSA_vs_NLO_ODE} (see also Fig.~\ref{fig:XODE_vs_SEOB_debug})}. In this example, we find a moderate error in $\alpha$ of $1.4\,{\rm rad}$. For BBHs with smaller $q=M_2/M_1$ and more negative spins, the error in the Euler angles can be more significant, leading to more significant mismatches in the waveforms~(for a more systematic exploration over parameter space, please see Sec.~\ref{sec:MM_par_space}). 

As a consequence of $t_{\rm gw} < t_{\rm pr}$ in the late inspiral stage, $\alpha$ plateaus to nearly a constant when it is \teja{viewed}{plotted} as a function of frequency. Indeed, precession becomes slow compared to the orbital decay, and different vectors do not have time to change significantly. Such behaviors can also be seen in, e.g., examples presented in Ref.~\cite{Chatziioannou:17}\teja{}{ though it was not emphasized}. Assuming we can extrapolate the PN dynamics, this motivates our choice of simply fixing the Euler angles after the merger, consistent with one of the options provided by Ref.~\cite{Gamba:22}. \teja{We leave more sophisticated treatments of the merger-ringdown phase (e.g., Ref.~\cite{Hamilton:21})  to future upgrades.}{More sophisticated treatment (e.g., Ref.~\cite{Hamilton:21}) in the merger-ringdown phase is left for future upgrades.}

\begin{figure}
  \centering
  \includegraphics[width=\columnwidth]{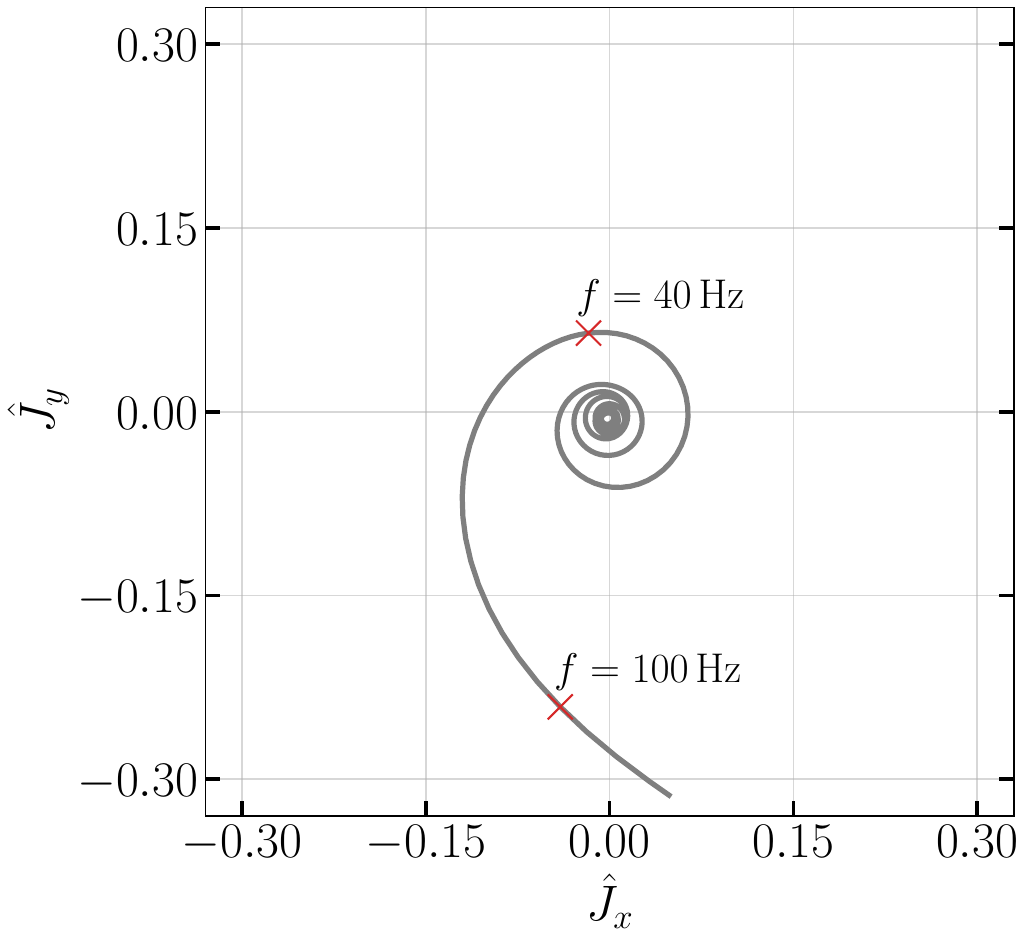}
\caption{Direction of the projection of the total angular momentum $\uvect{J}$ into the $x-y$ plane of the $J_0$ frame, for the system considered in Fig.~\ref{fig:timescale_comp} (specifically the one $\chi_{\rm eff}=-0.5$). The direction of $\uvect{J}$ is not a constant and its deviation from the origin increases as the orbit decays. The two red markers indicate the instants when $f=40\,{\rm Hz}$ and $100\,{\rm Hz}$, respectively. 
}
\label{fig:uJ_dir}
\end{figure}

Another assumption made by both the MSA and the NNLO construction is that $\uvect{J}$ is approximated as constant. 
A well-known example violating this assumption is BBHs undergoing transitional precession~\cite{Apostolatos:94}. This happens for BBHs with very asymmetric mass ratios ($q\ll 1$) and large, anti-aligned spins ($\chi_{1z}\sim -1$)\teja{. When constructing phenomenological frequency-domain approximants, the loss of accuracy in this regime is usually taken as an acceptable tradeoff}{, which are cases typically ignored in the construction of frequency-domain phenomenological approximants} as such systems occupy only a small fraction of the parameter space. 
However, in Fig.~\ref{fig:uJ_dir}, we show that even for a ``typical'' BBH, the assumption of constant $\uvect{J}$ can be violated to a significant extent. Here the system is the one we considered in Fig.~\ref{fig:timescale_comp} with $q=0.6$ and $\chi_{\rm eff}=-0.5$. As the inspiral proceeds, the deviation of $\uvect{J}$ from its initial value (the origin in Fig.~\ref{fig:uJ_dir}) increases. Moreover, because $t_{\rm pr} > t_{\rm gw}$ when $f\gtrsim 40\,{\rm Hz}$, $\uvect{J}$ does not have enough time to finish a complete precession cycle. Therefore, even the averaged orientation of $\uvect{J}$ when $f\gtrsim 40\,{\rm Hz}$ (where about half of the SNR$^2$ is expected; bottom panel of Fig.~\ref{fig:timescale_comp}) significantly deviates from $\uvect{J}_0$.

\subsection{Efficient precession evolution}
\label{sec:fast_prec}

Given the limitations described in Sec.~\ref{sec:limitations_p_avg}, we use the numerical solutions to the N4LO precession equations to compute the Euler angle. However, we would like to maintain the efficiency of the waveform construction. \teja{In this subsection, we present a technique to speed up the evaluation of the numerical solutions.}{We thus present in this subsection a technique to efficiently obtain the numerical solutions.}  

While both the NNLO and the MSA angles have their limitations, they can nonetheless serve as a good baseline  
solution that captures most of the dynamics. If we just numerically solve deviations from the analytical approximation, the deviation will typically be small and slowly varying and thus can be accurately integrated with only a few steps. This is the key idea we employ in our waveform construction. 

In practice, we find that a simple coordinate rotation 
can already sufficiently accelerate the precession evaluation. Specifically,
we define
\begin{align}
    \begin{pmatrix}
        \uLN^{(1)}\\
        \uvect{S}_1^{(1)} \\
        \uvect{S}_2^{(1)}
    \end{pmatrix}
    = \boldsymbol{R}_{J_0}\left[-\alpha^{(0)}\right]
    \begin{pmatrix}
        \uLN\\
        \uvect{S}_1 \\
        \uvect{S}_2
    \end{pmatrix},
    \label{eq:transformation}
\end{align}
where $\boldsymbol{R}_{J_0}\left[-\alpha^{(0)}\right]$ denotes a rotation around $\uvect{J}_0$ by an angle $-\alpha^{(0)}$. 
Here $\alpha^{(0)}$ is an analytical estimation of the precession angle. Here we adopt the value from a single-spin PN expansion at NLO~\cite{Pratten:21},
\begin{equation}
    \alpha^{(0)} = \sum_{k=-3}^{-1} A_k \left(M\omega\right)^k + \alpha_0^{(0)},
    \label{eq:al_0}
\end{equation}
where 
\begin{align}
    A_{-3} =& \frac{5 \delta }{64 M_1} - \frac{35}{192},  \\
    A_{-2} =& \frac{5 \chi_{\rm eff} M_1 \left(3\delta - 7 M_1\right) }{128\eta}, \\
    A_{-1} =& -\frac{5515}{3072} 
        + \eta \left(-\frac{515}{384} + \frac{175\delta}{256 M_1}- \frac{15 \delta^2}{256 M_1^2}\right) \nonumber \\
    &+\frac{4555\delta }{7168 M_1} + \frac{5 \chi_p^2 M_1^3\left(3 \delta - 7 M_1\right) }{128\eta^2},
\end{align}
and $\alpha_0^{(0)}$ is chosen such that at $f_{\rm ref}$, $\alpha = \alpha^{(0)}$. 
We have defined $\delta = (M_1-M_2)/M$. 
The baseline estimation for $\uLN^{(1)}$, etc., can be made more accurate if we use a more accurate approximation for $\alpha^{(0)}$ and/or account for two-spin effects, both are available in the MSA construction~\cite{Chatziioannou:17, Kesden:15}. 
We use Eq.~(\ref{eq:al_0}) here because it has especially simple forms for $\alpha^{(0)}$ and $d\alpha^{(0)}/d\omega$ as functions of $\omega$, and the acceleration it provides is sufficient for our purpose (see Sec.~\ref{sec:validation}). 
The evolution of the rotated vectors satisfies
\begin{multline}
    \frac{d}{d\omega}
    \begin{pmatrix}
        \uvect{L}_{\rm N}^{(1)}\\
        \uvect{S}_1^{(1)} \\
        \uvect{S}_2^{(1)}
    \end{pmatrix}\\
    = \boldsymbol{R}_{J_0}\left[-\alpha^{(0)}\right]
    \left[
    \frac{d}{d\omega}
    \begin{pmatrix}
        \uvect{L}_{\rm N}\\
        \uvect{S}_1 \\
        \uvect{S}_2
    \end{pmatrix}
    -\frac{d{\alpha}^{(0)}}{d\omega}
    \uvect{J}_0 \times 
    \begin{pmatrix}
        \uLN\\
        \uvect{S}_1 \\
        \uvect{S}_2
    \end{pmatrix}
    \right].
\end{multline}
This way we remove fast oscillations in $\uLN$, making $\uLN^{(1)}$ slowly varying. 
Furthermore, we can use the rotational invariance of the cross product, $\vect{R}(\vect{a} \times \vect{b}) = (\vect{R}\vect{a}) \times (\vect{R}\vect{b})$, to directly write the precession equations in the rotated frame in terms of $\left[ \duvect{L}_{\rm N}^{(1)}, \duvect{S}_1^{(1)}, \duvect{S}_2^{(1)}\right]$. 

\begin{figure}
  \centering
  \includegraphics[width=\columnwidth]{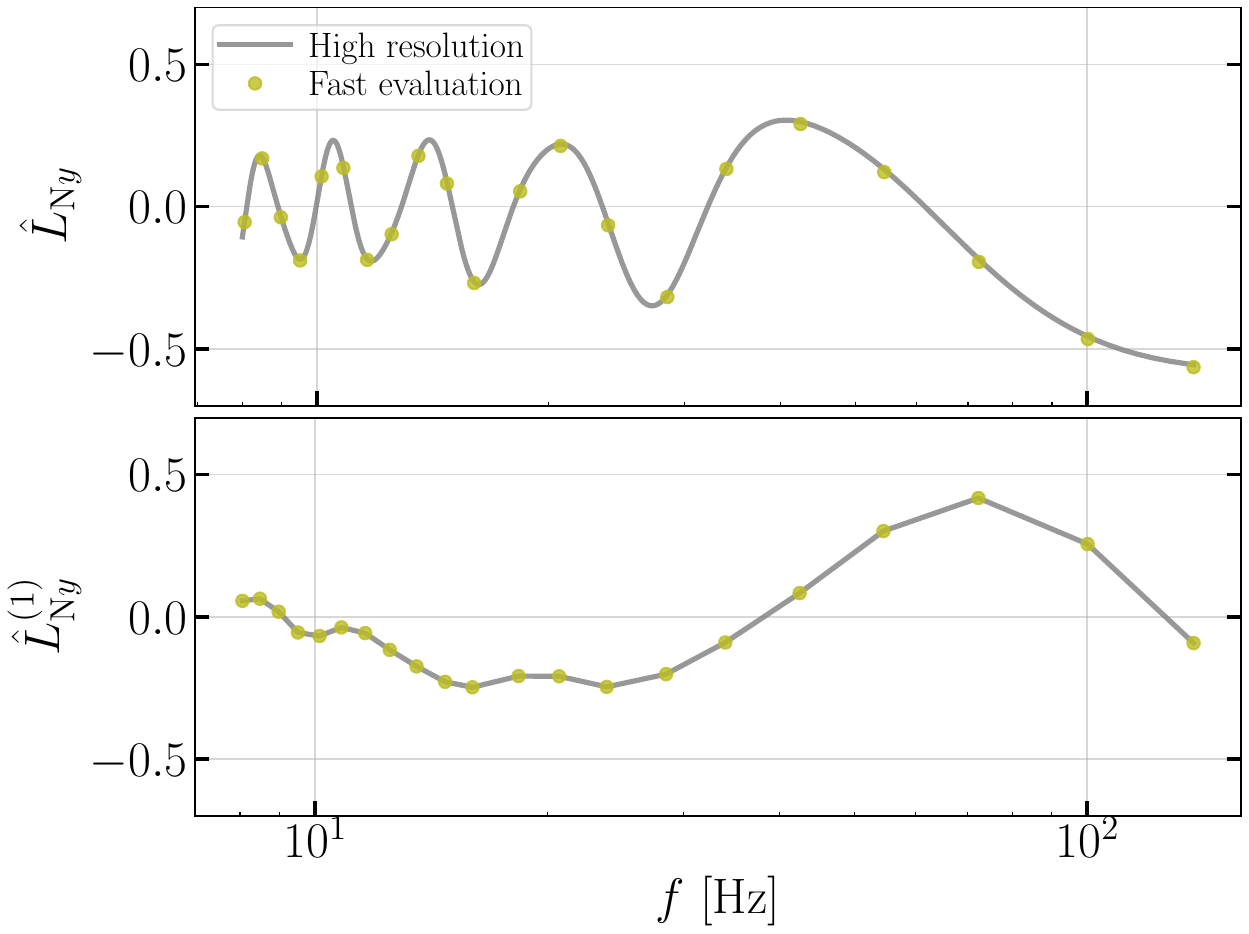}
\caption{{\em Top:} Evolution of the y-component of the direction of the orbital angular momentum, $\hat{L}_{{\rm N} y}$, in the inertial $J_0$ frame for the system in Fig.~\ref{fig:timescale_comp} with $\chieff=-0.5$. {\em Bottom:} Evolution of the rotated version $\hat{L}_{{\rm N} y}^{(1)}$, defined in Eq.~\eqref{eq:transformation}. Each dot indicates a step in the ODE. As $\hat{L}_{{\rm N} y}^{(1)}$ varies less compared to $\hat{L}_{{\rm N} y}$, it can be accurately integrated with only a few steps, which speeds its evaluation up. We thus solve $\hat{L}_{{\rm N}}$ numerically and then rotate it back to the inertial frame. }
\label{fig:fast_evol_demo}
\end{figure}

In Fig.~\ref{fig:fast_evol_demo} we present an example demonstrating how the evaluation of the precession dynamics is accelerated by the transformation given by Eq.~(\ref{eq:transformation}). We show in the top panel the $y$ component of $\uLN$. It exhibits multiple precession cycles and $\hat{L}_{Ny}$ oscillates between $-0.3$ and $0.3$. In comparison, $\hat{L}_{Ny}^{(1)}$ is a smoother curve that varies slower, as shown in the bottom panel. 
In this example, the transformation allows us to integrate the ODE in $\sim 20$ steps using a standard explicit Runge-Kutta method where the error is controlled assuming the accuracy of the fourth-order method, but steps are taken using the fifth-order accurate formula~\cite{Dormand:80}.
Each numerical step is represented by an olive dot in the plot. The bottom panel shows the result returned directly by the integrator in terms of $\hat{L}_{N}^{(1)}$ and in the top panel, the dots are obtained by inverting Eq.~(\ref{eq:transformation}) to reconstruct $\hat{L}_{N}$. 
Further reduction in the number of steps is limited by the need to resolve the spin-spin interaction, which is not captured by the simple coordinate transformation in Eq.~(\ref{eq:transformation}) but can be accounted for by utilizing the MSA results~\cite{Chatziioannou:17}. 
In comparison, direct integration of the precession equations requires $\sim 40$ steps to achieve the same numerical accuracy. 
We observe a more significant reduction in the number of evaluation steps when we require a higher numerical accuracy (i.e., smaller tolerance on the numerical errors).  
The evolution of $\hat{L}_{N}$  provides us with the Euler angles as functions of the orbital frequency $\omega$, which we then interpolate during the twisting-up process [Eqs.~(\ref{eq:hlm_L_2_hp_J}) and (\ref{eq:hlm_L_2_hc_J})]. 

In the special case where the BBH experiences transitional precession~\cite{Apostolatos:94}, the transformation in Eq.~(\ref{eq:transformation}) \teja{should no longer be able to take out the leading behavior, and thus we will not be able to significantly accelerate the solution of the ODE.}{would degrade in its ability to accelerate the numerical ODE evaluation.} Nevertheless, even in this special case, we can still obtain the exact evolution of $\uLN^{(1)}$ and hence $\uLN$. Therefore, \xode\ can describe BBHs with generic spin configurations, including those leading to transitional precessions.

\section{Recalibration of coprecessing modes}
\label{sec:cp_modes}

\begin{figure}
  \centering
  \includegraphics[width=0.95\columnwidth]{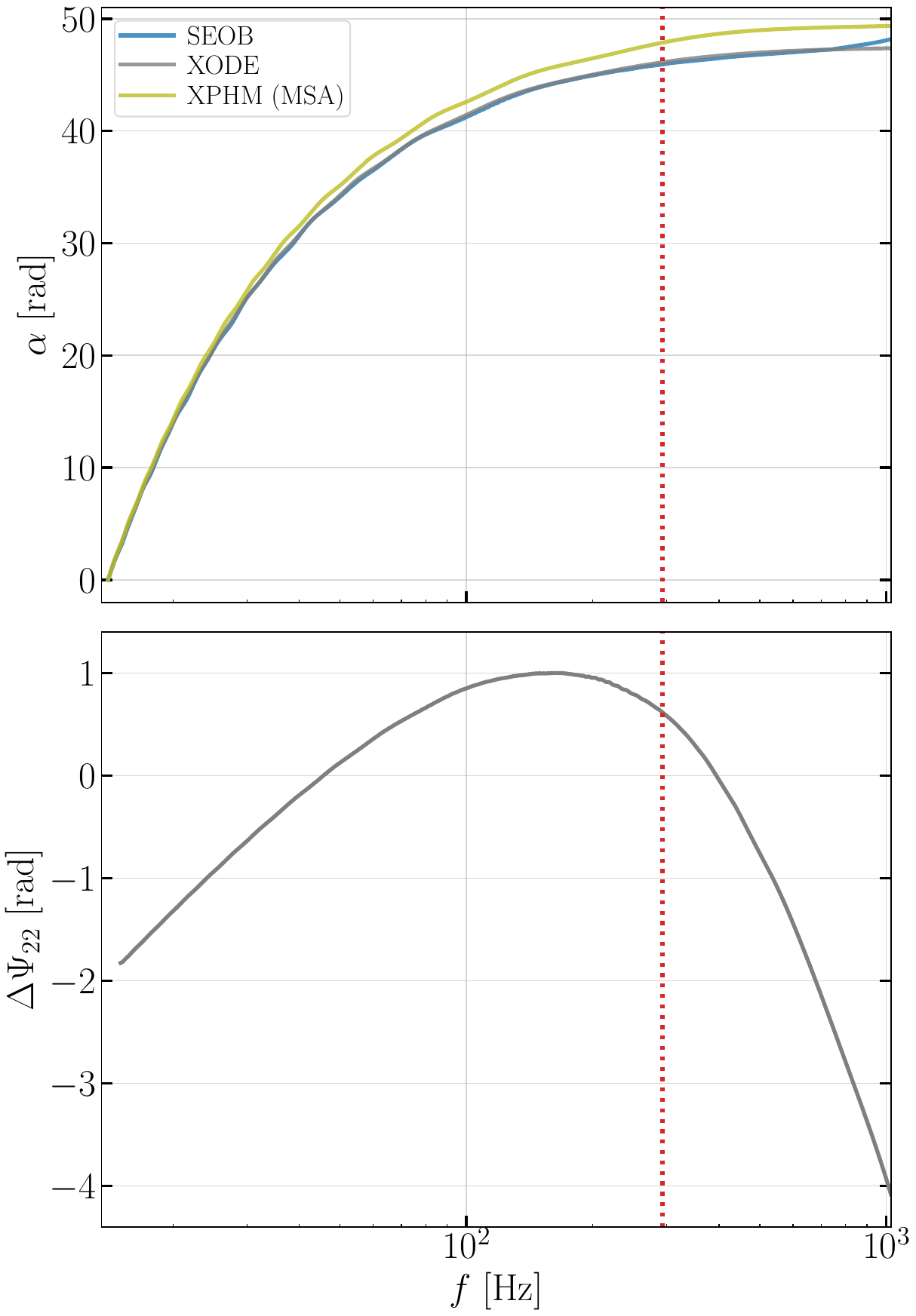}
\caption{{\em Top:} Comparison of the Euler angle $\alpha$, defined in Eq.~\eqref{eq:alphadef}, computed by different approximants (blue for \seob, gray for \xode, and olive for \xphm\ using the MSA angles). {\em Bottom:} phase difference of the coprecessing $(2, -2)$ mode (in the frequency domain) between \seob\ and \xphm. The vertical line indicates the location where the binary separation is $6M$. 
}
\label{fig:XODE_vs_SEOB_debug}
\end{figure}

Besides the Euler angles, GW modes in the coprecessing $L$ frame, $\tilde{h}_{\ell m'}^L$, are the other key ingredient for constructing the final waveform. It is commonly assumed that the coprecessing modes can be approximated by their non-precessing values up to an update of the final spin~\cite{Schmidt:11, Schmidt:12,  Hannam:14, Pratten:21}.
However, it is also well known that the combinations of $(\chi_{1z}, \chi_{2z})$ entering the PN frequency evolution~\cite{Cutler:94} are not conserved by the precession dynamics. \teja{The time-domain \seob\ approximant tracks the evolution of $(\chi_{1z}, \chi_{2z})$ along with the precession dynamics~\cite{Ossokine:20}, but this is not incorporated into the frequency-domain \xphm ~approximant.}{The evolution of $(\chi_{1z}, \chi_{2z})$ is tracked by the time-domain approximant \seob\ together with the precession dynamics~\cite{Ossokine:20}, but not by \xphm.} The coupled evolution also enables \seob\ to more self-consistently estimate the size of the final spin~\cite{Ossokine:20}, whereas in \xphm\ the contribution from the in-plane spins are only estimated based on either $\chi_p$ or the precession-averaged magnitude of the total spin~\cite{Pratten:21}. 
When comparing $\tilde{h}_{\ell m'}^L$ computed from the two approximants, we notice that \teja{}{sometimes} the difference can be significant \teja{depending on the parameters}{}. \teja{We would like to incorporate this physics into our waveform construction. We achieve this by {\em recalibrating} the phases of the coprecessing waveforms to match those in \seob.}{Since more physics is incorporated in the construction of \seob\ waveforms, we describe here our recalibration of the coprecessing modes according to \seob.  }


\teja{Fig.~\ref{fig:XODE_vs_SEOB_debug} compares  \xode, \xphm, and \seob, for a BBH system with $(M_1, M_2)=(12\,M_\odot, 3\,M_\odot)$ and $(\chi_{1z}, \chi_{1p})=(\chi_{2z}, \chi_{2p})=(0.2, 0.6)$ defined at $f_{\rm ref}=14\,{\rm Hz}$.\footnote{For all the comparisons involving \xphm, its default behavior is used. In particular, we set \texttt{PrecVersion} to 223 which utilizes the MSA Euler angles. }}{To see the need for modification of the coprecessing modes explicitly, an example comparing \xode, \xphm, and \seob\ is  presented in Fig.~\ref{fig:XODE_vs_SEOB_debug}.\footnote{For all the comparisons involving \xphm, its default behavior is used. In particular, we set \texttt{PrecVersion} to 223 which utilizes the MSA Euler angles. } 
The system has $(M_1, M_2)=(12\,M_\odot, 3\,M_\odot)$ and $(\chi_{1z}, \chi_{1p})=(\chi_{2z}, \chi_{2p})=(0.2, 0.6)$ defined at $f_{\rm ref}=14\,{\rm Hz}$.} The vertical dotted line indicates the location when the binary separation is $6M$. 
The top panel compares again the Euler angle $\alpha$ computed by different approximants. We note that the \xode\ angle agrees well with the \seob\ result. The MSA angle employed by \xphm, however, overestimates $\alpha$ by a few radians \teja{as it was}{due to both the fact that it is} derived at a lower PN order \teja{, and due to the intrinsic limitations of the MSA construction}{and that the MSA construction has intrinsic limitations} (see Sec.~\ref{sec:limitations_p_avg} and Fig.~\ref{fig:MSA_vs_NLO_ODE}). 

Fixing the Euler angle alone does not improve the match between phenomenological models and \seob\ as we will see later in Sec.~\ref{sec:validation}. The reason is illustrated in the bottom panel of Fig.~\ref{fig:XODE_vs_SEOB_debug}. 
Here we define
\begin{equation}
    \Delta \Psi_{\ell,|m'|}(f)= \angle \tilde{h}^L_{\ell,|m'|}(f)\big|_{\rm SEOB} - \angle \tilde{h}^L_{\ell,|m'|}(f)\big|_{\rm XPHM},
\end{equation}
where we use the ``$\angle$'' symbol to denote the phase of a complex number. Thus $\Delta \Psi_{22}$ measures the phase difference between the coprecessing $(2, -2)$ modes generated by \seob\ and \xphm\ at a given frequency $f$. 
Note $\tilde{h}^L_{\ell,|m'|}\big|_{\rm SEOB}$ is obtained by Fourier transforming the time-domain modes computed by \seob. Following the \LAL~convention, only modes with $m'<0$ have support for $\tilde{h}^L$ for $f>0$.\footnote{Consistent with Refs.~\cite{Ossokine:20, Pratten:21}, we ignore asymmetries of the coprecessing modes~\cite{Pekowsky:13,Boyle:14} in the construction of \xode.} We can thus quote only the absolute value of $m'$ in the subscript without ambiguity. We have also removed the linear part of $\Delta \Psi$ (with a weight proportional to the SNR squared)\teja{, which corresponds to maximizing the match of the two waveforms over the choice of overall time and constant phase shifts between the two waveforms \cite{PhysRevD.99.123022}.}{ corresponding to an overall time and constant phase shift of the waveform.} From the plot, the variation in $\Delta \Psi$ for the dominant $(2, -2)$ mode can be $3\,{\rm rad}$ just within the inspiral part, comparable to the error in the Euler angle. Such a significant deviation may lead to large mismatches in the waveforms, as we will see in Sec.~\ref{sec:validation}.

\begin{figure}
  \centering
  \includegraphics[width=0.95\columnwidth]{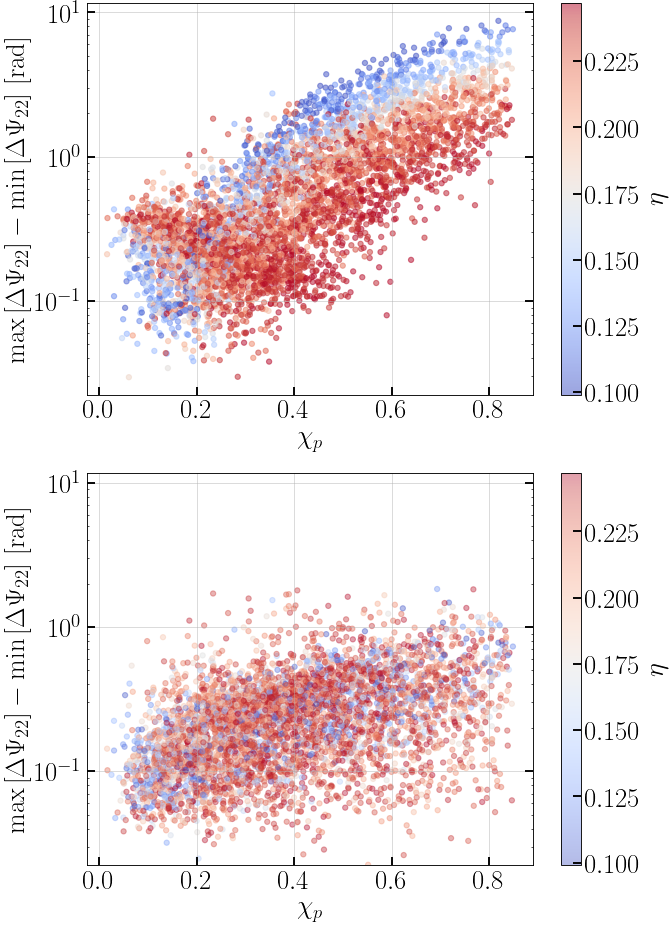}
\caption{Scatter plot showing the maximum difference of the coprecessing (2, -2) mode's phase mismatch between two approximants,
$\left(\max_f\left[\Delta \Psi_{22}\right] - \min_f\left[\Delta \Psi_{22}\right]\right)$ with $Mf \in [0.001, 0.022]$, versus different values of $\chi_p$. 
We further color code each point according to $\eta=q/(1+q)^2$. 
The top panel compares the difference between \seob\ and \xphm. The bottom panel shows the result after we perform the recalibration. 
}
\label{fig:coprec_ang_diff_vs_chip}
\end{figure}

More systematically, we randomly generate 5000 BBH systems and compute $\Delta \Psi_{22}$. We fix the total mass to be $M=15\,M_\odot$ but uniformly and independently sample the mass ratio $q\in [0.125, 0.8]$ and the magnitude of each spin $\chi_{1, (2)}\in [0.1, 0.85]$.  The orientation of each spin is sampled isotropically. The value of $\Delta \Psi$ for symmetric systems with small spins is small and therefore we exclude them in the recalibration. In the top panel of Fig.~\ref{fig:coprec_ang_diff_vs_chip}, we show a scatter plot for the variation in $\Delta \Psi_{22}$ in the inspiral part (approximated by a binary separation greater than $6M$) as a function of $\chi_p$. We further color-code the points according to $\eta$. We note that $\Delta \Psi_{22}$ can be especially large for large $\chi_p$ and small $\eta$. For those systems, approximating the coprecessing modes with their non-precessing value may not be sufficient. 

Nevertheless, we note the phase difference is in general a smooth function of frequency (bottom panel of Fig.~\ref{fig:XODE_vs_SEOB_debug}). Indeed, the coprecessing modes used by \xphm~should have already captured the major features in the phase evolution. We find that the residual phase difference  over the entire frequency range (i.e., including both inspiral and merger-ringdown parts) can be well fitted by a phenomenological model
\begin{equation}
    \Delta \Psi_{\ell, |m'|}(f) = \sum_{k=0}^{3} \lambda_{\ell, |m'|, k} \left[\ln \left(Mf\right)\right]^{(3-k)}.
    \label{eq:dPsi_phenom_fit}
\end{equation}
The constant part in $\Delta \Psi_{\ell, |m'|}(f)$ is determined by setting $\Delta \Psi_{\ell, |m'|}=0$ at $f=f_{\rm ref} (|m'|/2)$ and the linear-in-$f$ part has been removed before fitting Eq.~(\ref{eq:dPsi_phenom_fit}). We then fit the phenomenological parameters $\vect{\lambda} = [\lambda_{0}, \lambda_{1}, \lambda_{2}]$ (note we have dropped $\ell, |m'|$ in the subscript for conciseness) in terms of the physical parameters of the system
\begin{equation}
    \vect{\lambda} =  \vect{\xi} \vect{\Lambda},
    \label{eq:phs_2_phenom}
\end{equation}
where the physical parameters of the system which we 
empirically
determine to use are given by the vector $\vect{\xi}$
\begin{align}
    \vect{\xi} = [&
    1, \eta, \eta^2, \chi_{\rm eff}, \eta\chieff, \eta^2\chieff, \nonumber \\
    &\chi_p, \eta \chi_p, \eta^2 \chi_p, 
    \chieff \chi_p, \eta \chieff \chi_p, \nonumber \\
    &\chi_p^2, \eta \chi_p^2, \eta^2 \chi_p^2
    ].
    \label{eq:phys_par_vect}
\end{align}
In our code, we evaluate $\chi_p$ at a binary separation of $6M$, yet using $\chi_p$ at a different instance does not significantly impact the result. 
\teja{We evaluate the coprecessing waveforms for a set of random parameter values, and find the $14\times3$ coefficient matrix $\vect{\Lambda}$ that generates a set of phase differences via Eqs.~\eqref{eq:phs_2_phenom} and \eqref{eq:dPsi_phenom_fit} that are closest to the observed phase differences in a least-squared sense. Appendix~\ref{appx:Lambda} presents the numerical results for the coefficients.}{The $14\times3$ coefficient matrix $\vect{\Lambda}$ is found by finding the least-squares solution to Eq.~(\ref{eq:phs_2_phenom}) whose numerical results are presented in Appendix~\ref{appx:Lambda}.} 
The phase difference $\Delta \Psi_{22}$ after the recalibration is shown in the bottom panel of Fig.~\ref{fig:coprec_ang_diff_vs_chip}. The typical variation in $\Delta \Psi_{22}$ is now around $0.2-0.3\,{\rm rad}$, and for most of the systems it stays within $1\,{\rm rad}$. 
We can also include other information such as the individual $(\chi_{1p}, \chi_{2p})$ and the relative angle between $\vect{S}_1$ and $\vect{S}_2$ in $\vect{\xi}$, but we found that including them did not significantly improve the result. 
To further reduce the residual, we can, e.g., use the principal components of $\Delta \Psi_{\ell, |m'|}$ to replace the simple model of Eq.~(\ref{eq:dPsi_phenom_fit}), which we plan to explore in future upgrades. 
In the current version, we perform the fit for $(\ell, m)=(2, -2)$, $(3, -3)$, and $(2, -1)$ coprecessing modes. 

To summarize, the steps in the construction of our waveform approximation, \xode, are as follows:
\begin{enumerate}
    \item \textbf{Generation of coprecessing modes:} We use the functionality provided in \LAL~\cite{LAL:18, SWIGLAL:20} to generate the same modes $\tilde{h}_{\ell m'}^{L}$ of the waveform in the coprecessing frame that \xphm\ uses. 
    The coprecessing modes available in \xphm\ include $(\ell, |m'|)=(2, 2), (2, 1), (3, 3), (3, 2), (4, 4)$.
    We then compute an additional {\em calibration} phase $\Delta \Psi_{\ell, |m'|}$ over the entire frequency range according to Eq.~(\ref{eq:dPsi_phenom_fit}), to be added to the coprecessing modes so that they match results from \seob\ for strongly precessing BBHs. In the current implementation, we have fitting functions for the calibration phase for the $(2, 2)$, $(2, 1)$, and $(3,3)$ modes. We multiply $\Delta \Psi_{\ell, |m'|}$ by an overall factor $\mathcal{W}=20\chi_p$ if $\chi_p<0.05$ and $\mathcal{W}=1$ if $\chi_p\geq 0.05$ so that \xode\ falls back to \texttt{IMRPhenomXHM}~\cite{GarciaQuiros:20} in the non-precessing limit. 
    
    We emphasize that by adding $\Delta \Psi_{\ell, |m'|}$, we are making the assumption that the \seob\ construction provides a more accurate description of the physical system than \xphm. This is physically motivated as \seob\ accounts for more physics effects such as the time-dependent evolution of $\chi_{1z(2z)}$ as well as a more self-consistent estimation of the final spin. On the other hand, \seob\ does not perfectly agree with numerical relativity, and its coprecessing modes are not specifically calibrated to NR (though the final waveform from \seob\ has been validated against NR~\cite{Ossokine:20}). Hence, the user has the choice of whether to include the recalibration phase $\Delta \Psi_{\ell, |m'|}$ or not. The default behavior is to include $\Delta \Psi_{\ell, |m'|}$. 

    \item \textbf{Solution of the Euler angles:} We numerically solve the N4LO precession equations with the acceleration technique described in Sec.~\ref{sec:fast_prec} to obtain the Euler angles $(\alpha, \beta, \epsilon)$ as functions of orbital frequency $\omega$. 

    \item \textbf{Twisting up:} Lastly, the coprecessing modes are twisted up according to Eqs.~(\ref{eq:hlm_L_2_hp_J}) and (\ref{eq:hlm_L_2_hc_J}) under the stationary phase approximation to transform to the inertial frame. In the limit in which the BBH is non-precessing (with $\chi_p<10^{-6}$), our code will call \texttt{IMRPhenomXHM}~\cite{GarciaQuiros:20}. 
    We have also verified that \xode\ is consistent with \texttt{IMRPhenomXHM} for small but finite $\chi_p$, even for systems with asymmetric mass ratios and strongly anti-aligned spins. 
\end{enumerate}

Our code is currently written in Python with standard \texttt{NumPy}~\cite{Harris:20} and \texttt{SciPy}~\cite{Virtanen:20} packages as well as \texttt{python-lalsimulation} (\cite{LAL:18, SWIGLAL:20}; version 5.1.0). The computational efficiency is further accelerated by \texttt{Numba}~\cite{Lam:15}. \xode\ is publicly available at \url{https://github.com/hangyu45/IMRPhenomXODE}. 

\section{Model performance and validation}
\label{sec:validation}

\begin{figure}
  \centering
  \includegraphics[width=0.95\columnwidth]{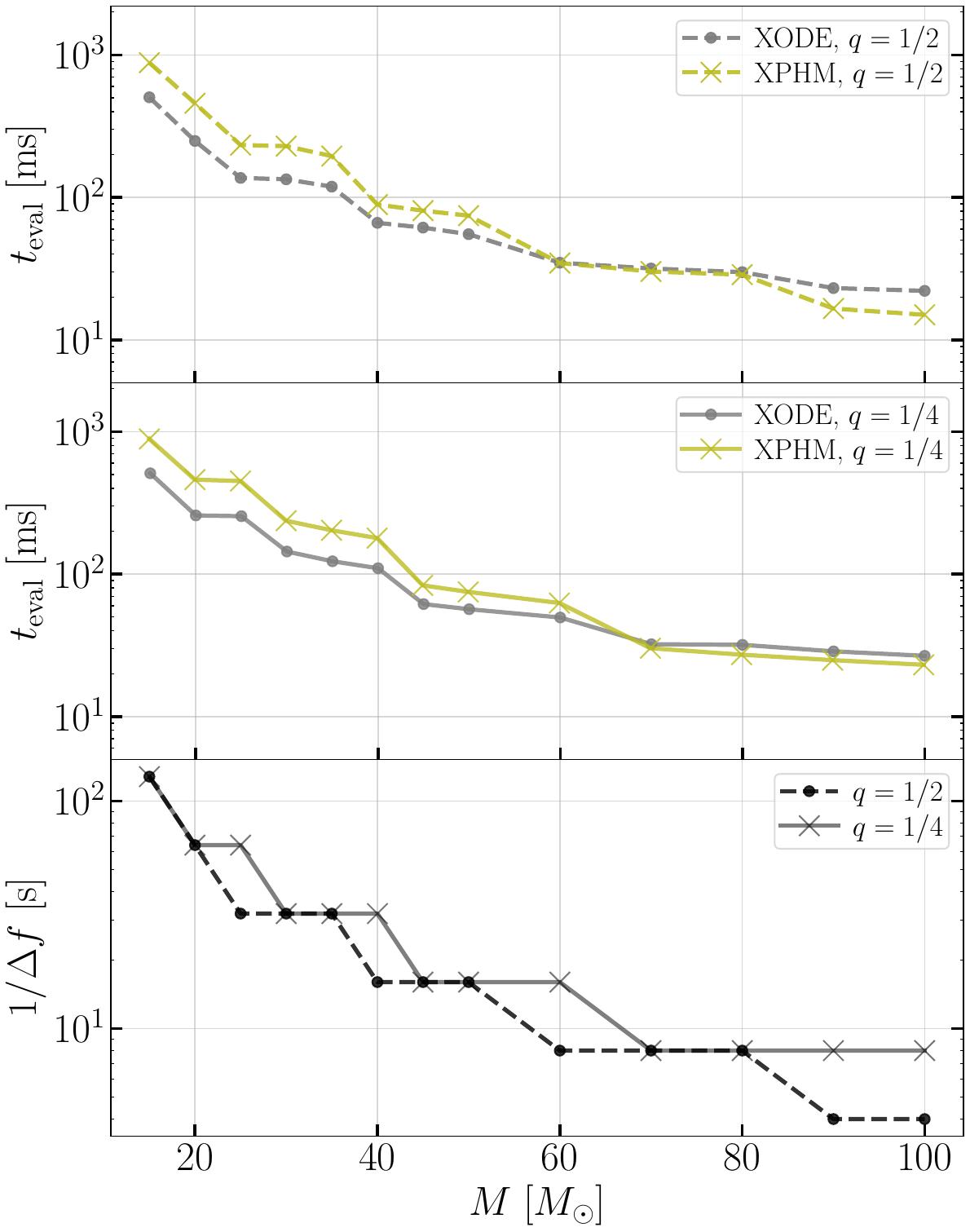}
\caption{The top two panels show the evaluation time per waveform generation, $t_{\rm eval}$, as a function of the total mass $M$ of the system. The total mass determines the resolution and the upper limit of the frequency grid. In the top two panels, we use the color (gray, olive) and the (dot, cross) marker to represent (\xode, \xphm). The (dashed, solid) lines are used to represent $q=(1/2, 1/4)$. The initial GW frequency is fixed at 10\,Hz. The bottom panel shows the inverse of the frequency resolution, which is related to the signal duration via Eq.~(\ref{eq:resolution}). When evaluating the waveform over a large frequency grid, \xode\ can be generated even more efficiently than \xphm. As the size of the frequency grid decreases, \xode\ hits a bottleneck of $t_{\rm eval}\sim 20\,{\rm ms}$ set by the requirement to numerically solve the precession ODEs. 
}
\label{fig:evaluation_time}
\end{figure}

In this Section, we assess the performance of \xode.

As a phenomenological frequency-domain approximant, \xode\ can be efficiently generated as shown in Fig.~\ref{fig:evaluation_time}. Here we report the evaluation time per waveform of \xode\ on a personal laptop (specifically, a MacBook Pro with a 2.6 GHz 6-Core Intel Core i7 processor) and compare it with \xphm\ (using default parameters), which is one of the most efficient waveform approximants in the literature. The system we consider has a spin configuration $(\chi_{1x}, \chi_{1y}, \chi_{1z}) = (0, 0.6, 0.2),\ (\chi_{2x}, \chi_{2y}, \chi_{2z})=(0.6, 0, 0.2)$ and always starts from a GW frequency of $f_{\rm low}=10$\,Hz. Because we include the $\ell=|m'|=4$ mode in the coprecessing $L$ frame, this means the precession ODE needs to be integrated starting from $\omega = 5\pi\,{\rm rad\,s^{-1}}$. Two different mass ratios are considered, $q=1/2$ (top panel) and $q=1/4$ (middle panel). We quote the average generation time, $t_{\rm eval}$, as a function of the total mass, which affects the evaluation time for two reasons. Firstly, we set the frequency resolution $\Delta f$ based on the duration of the waveform, as
\begin{equation}
    \log_{2}\left(\frac{1}{\Delta f}\right)= {\rm ceil}\left[\log_2(t_{\rm dur})\right], 
    \label{eq:resolution}
\end{equation}
and $t_{\rm dur}$ is an estimation of the merger time when the dominant $l=|m'|=2$ mode enters the frequency band, 
\begin{equation}
    \frac{t_{\rm dur}}{M} =\frac{5}{256\eta (\pi M f_{\rm low})^{8/3}}\simeq \frac{3}{8}\frac{t_{\rm gw}}{M}. 
\end{equation}
Secondly, the upper limit of the frequency vector is set to $f_{\rm up}={\rm min}[0.3/M, 2048\,{\rm Hz}]$. 
The frequency vector we use to generate the waveform then spans from $f_{\rm low}$ to $f_{\rm up}$ with a uniform spacing of $\Delta f$.\footnote{Note the choice of the frequency vector here is to account for the duration of the signal in the time domain. In practice, \xode\ can compute the waveform on an arbitrary frequency grid with $f>0$ and the grid can be unevenly sampled.} 

Interestingly, we note that \xode\ can outperform \xphm\ in terms of computational efficiency when the frequency grid has a large size (corresponding to long duration in the time domain and consequently high resolution in the frequency domain). When $M\lesssim 60\,M_\odot$, generating the coprecessing modes is the major computational cost in \xode.\footnote{For interested readers, a breakdown of the computational cost of \xode\ is provided in the GitHub repository.}
Since we use the same coprecessing mode as \xphm\ (with additional phase calibration as described in Sec.~\ref{sec:cp_modes}), the acceleration compared to \xphm\ is thus likely due to our efficient computation of the Euler angles (Sec.~\ref{sec:fast_prec}) and potentially a more optimized code structure for twisting up. As the size of the frequency grid decreases, \xode\ eventually hits a bottleneck set by numerically solving the ODEs and its evaluation becomes slower than \xphm.  We note that some efficient data analysis techniques~\cite{Roulet:22, Wong:23, Edwards:23} utilizing heterodyning or relative binning~\cite{Cornish:10, Cardoso:21, Zackay:18} do not require the full resolution but only a sparsely chosen subset of frequency bins. When those techniques are used, \xphm\ is still the most efficient waveform generating routine, and \xode\ is slower than \xphm\ by a factor of 2 to 3, but is still efficient with a typical evaluation time of $\sim 20$\,ms. 


\teja{We now proceed to test the accuracy of \xode\ over a range of parameter space}{More important is the accuracy of \xode}. Ideally, we would like to validate \xode\ with NR simulations~\cite{Boyle:19} or NR surrogates~\cite{Varma:19}. However, existing NR waveforms are too short to capture the rich precession dynamics happening mainly in the early inspiral part (Sec.~\ref{sec:limitations_p_avg}). This point will be made more explicit when we discuss Figs.~\ref{fig:eg_qq_25_chiz_n7_chip_3_vs_NR} and \ref{fig:eg_qq_25_chiz_n7_chip_3_vs_SEOB}. Therefore, we will use \seob~\cite{Ossokine:20} as our main point of comparison with the caveat that \seob\ itself is not yet a perfect description of NR.  

Following the literature, we quantify the accuracy of \xode\ by considering its match (\cite{Lindblom:08}, also known as the fitting factor) with a reference waveform (here we use \seob), 
\begin{equation}
    \mathcal{M}(h_1, h_2) = \max_{t_c, \phi_c}\frac{\langle h_1, h_2 \rangle}{\sqrt {\langle h_1, h_1 \rangle \langle h_2, h_2 \rangle}},
\end{equation}
where $(t_c, \phi_c)$ are time and phase shifts of $h_2$, and the inner product is defined as 
\begin{equation}
    \langle h_1, h_2 \rangle = 4 {\rm Re}\left[\int \frac{{h}_1(f) {h}^\ast_2(f)}{S_n (f)} df\right]. 
\end{equation}
Here $S_n (f)$ is the one-sided power spectral density of the detector noise, for which we use the aLIGO design sensitivity~\cite{Barsotti:18}. 

The squared SNR of a signal $h_1$ is given by $\rho^2 = \langle h_1, h_1 \rangle$. \teja{We can derive the requirement for the match by demanding that the deviations in the likelihood surface for events are small; this requirement shows that}{One can further show that} a mismatch \teja{(the complement of the match)}{} is significant if 
\begin{equation}
    1-\mathcal{M} \gtrsim \frac{1}{2\rho^2}.
\end{equation}
\teja{Given a typical level of mismatches for our waveform approximant and the `ground truth', this condition constrains the point up to which we can trust inference results \cite{PhysRevResearch.2.023151}.}{}

Note that a signal can be written in terms of its two polarizations as~\cite{Harry:16} 
\begin{equation}
    h = K \left( h_+ \cos \kappa + h_\times \sin \kappa\right),
\end{equation}
where $K$ is an overall amplitude that will be canceled out in $\mathcal{M}$ and $\kappa$ is an angle related to the antenna response 
When computing $\mathcal{M}$, we analytically optimize over $\kappa$ following Ref.~\cite{Harry:16}. In addition, we numerically optimize over the reference phase $\phi_{\rm ref}$ and rigid rotations of the in-plane spins at $f_{\rm ref}$ to be consistent with Ref.~\cite{Pratten:21}. 

\subsection{Specific examples}
\label{sec:MM_eg}

\begin{figure}
  \centering
  \includegraphics[width=0.95\columnwidth]{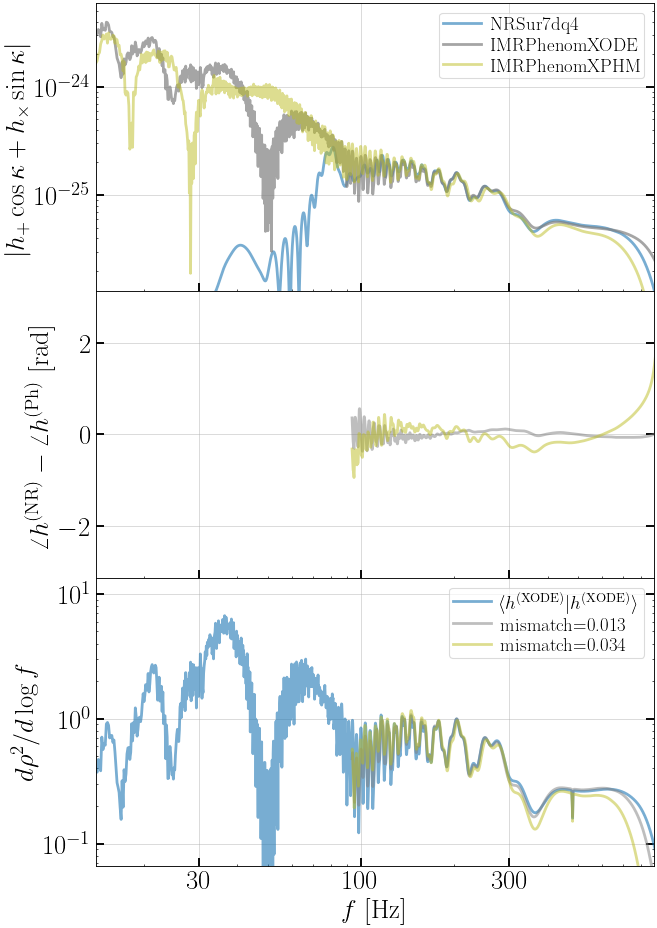}
\caption{Waveform comparison between \texttt{NRSur7dq4} (blue lines), \xode\ (gray lines), and \xphm\ (olive). From top to bottom, we show respectively the magnitude of the waveform, the phase difference between the NR surrogate and the phenomenological approximant, and the density of the SNR squared or match. The system considered here has $M=15\,M_\odot$, $q=1/4$, and $(\chi_{1z}, \chi_{1p})=(\chi_{2z}, \chi_{2p})=(-0.7, 0.3)$. }
\label{fig:eg_qq_25_chiz_n7_chip_3_vs_NR}
\end{figure}

\begin{figure}
  \centering
  \includegraphics[width=0.95\columnwidth]{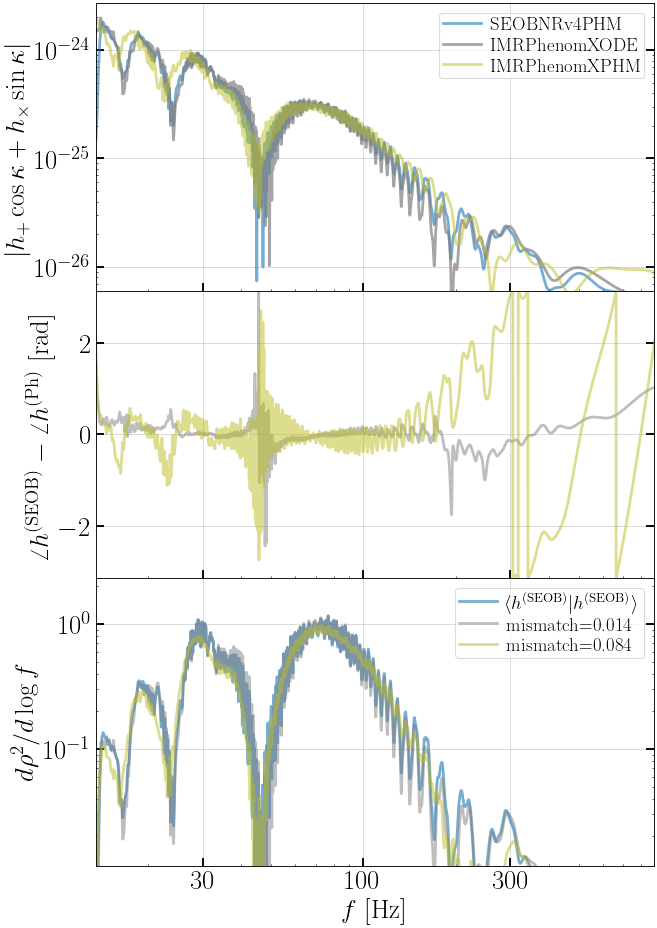}
\caption{Similar to Fig.~\ref{fig:eg_qq_25_chiz_n7_chip_3_vs_NR} but with \texttt{NRSur7dq4} replaced by \seob\, so that the reference waveform can be generated from an arbitrarily low frequency. The system has the same intrinsic parameters as the one shown in Fig.~\ref{fig:eg_qq_25_chiz_n7_chip_3_vs_NR} but with different $\phi_{\rm ref}$. }
\label{fig:eg_qq_25_chiz_n7_chip_3_vs_SEOB}
\end{figure}

\begin{figure}
  \centering
  \includegraphics[width=0.95\columnwidth]{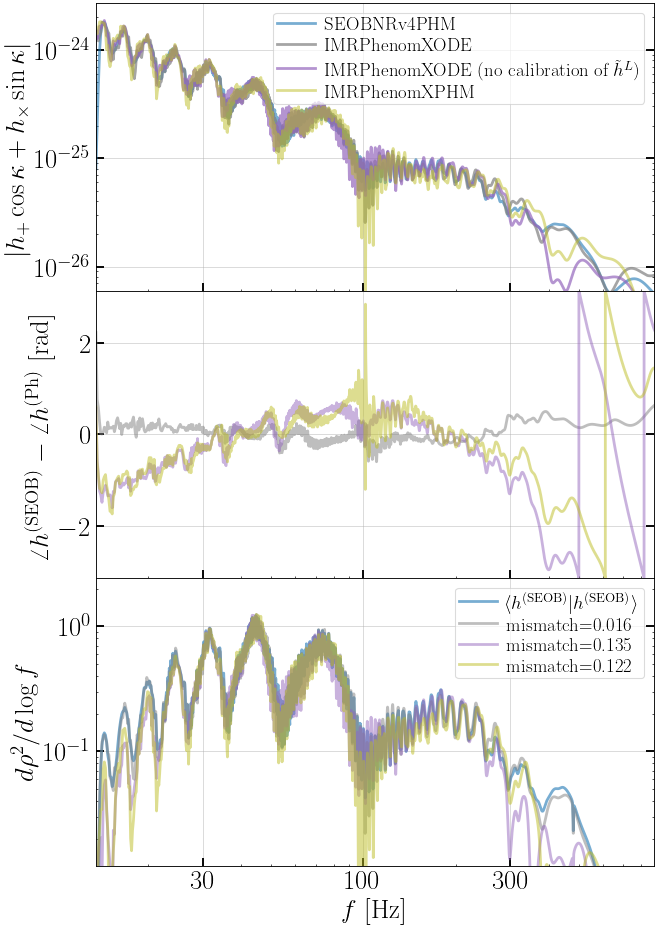}
\caption{Similar to Fig.~\ref{fig:eg_qq_25_chiz_n7_chip_3_vs_SEOB} but with the spins changed to $(\chi_{1z}, \chi_{1p})=(\chi_{2z}, \chi_{2p})=(0.2, 0.6)$ (same as the one considered in Fig.~\ref{fig:XODE_vs_SEOB_debug}). The new purple curves are generated by turning off the recalibration of coprecessing modes that is defaulted to be on in \xode. 
Without the coprecessing-mode recalibration, there can be a large mismatch between \xode\ and \seob\ despite the Euler angles being the same (see Fig.~\ref{fig:XODE_vs_SEOB_debug}). Adding back the recalibration (gray curves) can reduce significantly the mismatch.}
\label{fig:eg_qq_25_chiz_p2_chip_6_vs_SEOB}
\end{figure}

\begin{figure}
  \centering
  \includegraphics[width=0.95\columnwidth]{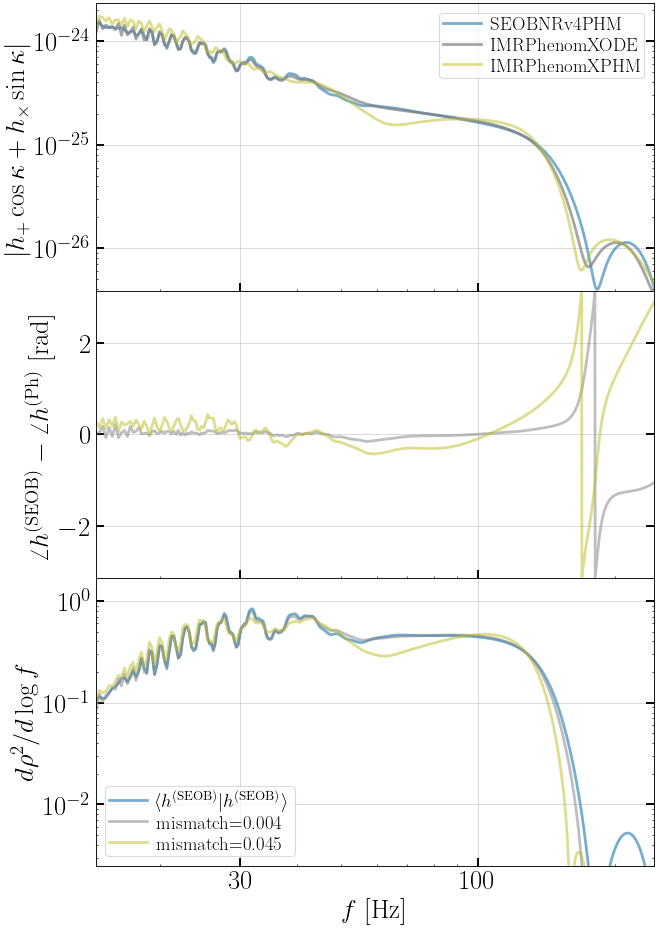}
\caption{An example demonstrating that \xode\ can improve the match against \seob\ even for massive BBHs. In this example, the BBH has $M=100\,M_\odot$, $q=1/4$, $(\chi_{1z}, \chi_{1p})=(-0.5, 0.4)$, and $(\chi_{2z}, \chi_{2p})=(-0.3, 0.7)$. }
\label{fig:eg_qq_25_chiz_n5_chip_4_vs_SEOB}
\end{figure}

We start our assessment of waveform accuracy by considering specific examples. 

\teja{Figure \ref{fig:eg_qq_25_chiz_n7_chip_3_vs_NR} illustrates why are unable to use NR and NR surrogates to validate our waveform model at this point in time}{We first elaborate on why NR and NR surrogates are not used to validate the waveforms by presenting Fig.~\ref{fig:eg_qq_25_chiz_n7_chip_3_vs_NR}}. Here we consider a system with $M=15\,M_\odot$ and $q=1/4$ with moderate precession $(\chi_{1z}, \chi_{1p}) = (\chi_{2z}, \chi_{2p}) = (-0.7, 0.3)$. We use $\iota=\pi/3$ and randomly pick $\kappa$, $\phi_{\rm ref}$, and the azimuthal orientations of $\chi_{1p}$ and $\chi_{2p}$.  In the top panel, we present the magnitude of the signal. The color (blue, gray, olive) represents waveform generated from (\texttt{NRSur7dq4}~\cite{Varma:19}, \xode, \xphm). The middle panel shows the phase difference between NR surrogate and phenomenological waveforms (only shown for frequencies where the NR surrogate is available). And lastly, the bottom panel shows the density of the SNR squared for the signal (blue) as well as the density of matches with NR 
(
i.e., it shows $d\langle h_1, h_2 \rangle / d\log f$;
we use the color gray and olive for \xode\ and \xphm, respectively). 
Note that we have normalized each waveform so that they all have $\rho=1$ in the band where  \texttt{NRSur7dq4} is defined ($f\gtrsim 90\,{\rm Hz}$). 

In this example, the \texttt{NRSur7dq4} waveform is available only for $f\gtrsim 90\,{\rm Hz}$, which contributes to only a small fraction of the total SNR of the system as shown in the bottom panel. Moreover, as we argued in Sec.~\ref{sec:limitations_p_avg}, at frequencies where NR is available, we have $t_{\rm gw} < t_{\rm pr}$ and the Euler angle $\alpha$ evolves only a small amount (see Figs.~\ref{fig:MSA_vs_NLO_ODE} and \ref{fig:XODE_vs_SEOB_debug}). A rigid rotation of the system can largely compensate for deviations in the evolution of $\alpha$. Therefore, both \xode\ and \xphm\ can provide decent matches to the NR waveform within this frequency range. However, as we extend down to lower frequencies, \xode\ and \xphm\ show significant discrepancies
as obvious in the top panel due to the different Euler angles used in the waveform construction.

Given the limitations of NR and NR surrogates, we instead use \seob~\cite{Ossokine:20} to validate our approximant. 
\seob~itself has been extensively validated against NR simulations and its mismatch with NR is below 3\% for 94\% of the systems. Only when $q\lesssim 1/4$ and $\chi_p \gtrsim 0.6$ does \seob~start to lose accuracy, and even in the worst case its mismatch with NR is within 10\%~\cite{Ossokine:20}.  
The comparison between \seob, \xode, and \xphm\ is shown in Fig.~\ref{fig:eg_qq_25_chiz_n7_chip_3_vs_SEOB}. The system considered here is similar to the one considered in Fig.~\ref{fig:eg_qq_25_chiz_n7_chip_3_vs_NR} but with a different $\phi_{\rm ref}$. We normalize the signal such that $\rho=1$ when integrated over the entire frequency band. When the early inspiral part is included, \xphm\ does not agree well with \seob\ for the entire range of frequencies as the MSA angles do not track the precession dynamics sufficiently accurately. As a result, the mismatch between \xphm\ and \seob\ can be 8.4\%, which can lead to large systematic errors in data analysis. In comparison, \xode\ shows a much better agreement with \seob\ with a mismatch of 1.4\% due to its more accurate computation of the Euler angles. 

\teja{Figure \ref{fig:eg_qq_25_chiz_p2_chip_6_vs_SEOB} demonstrates the necessity of recalibrating the coprecessing modes, which is the second key ingredient in our waveform model}{Besides improving the precession dynamics, recalibrating the coprecessing modes is another key ingredient whose necessity we demonstrate in Fig.~\ref{fig:eg_qq_25_chiz_p2_chip_6_vs_SEOB}}. The system we consider still has $M=15\,M_\odot$ and $q=1/4$, yet it experiences more precession as its spins are $(\chi_{1z}, \chi_{1p}) = (\chi_{2z}, \chi_{2p}) = (0.2, 0.6)$. This is the same BBH \teja{as the one}{} considered in Fig.~\ref{fig:XODE_vs_SEOB_debug}, where we showed the agreement between the Euler angle $\alpha$ that we numerically obtained and the one used in \seob. However, fixing the Euler angles alone while using the same coprecessing modes as used by \xphm\ leads to a waveform approximant (purple curves in Fig.~\ref{fig:eg_qq_25_chiz_p2_chip_6_vs_SEOB}) that still disagrees with \seob\ significantly with a mismatch of $13.5\%$. In fact, the mismatch is greater than using \xphm\ as the errors in the MSA angles partially cancel with the errors in the coprecessing modes. On the other hand, if we simultaneously fix the Euler angles and the phases of coprecessing $(2,-2)$, $(2,-1)$, and $(3,-3)$ modes (gray lines in Fig.~\ref{fig:eg_qq_25_chiz_p2_chip_6_vs_SEOB}), the agreement with \seob\ can be improved to $98.4\%$.

So far we have focused on light BBHs with $M=15\,M_\odot$ whose SNR  comes predominantly from the inspiral part. We would expect the improvement in our waveform construction to be most significant for such a system. 
For completeness, we consider also an example in Fig.~\ref{fig:eg_qq_25_chiz_n5_chip_4_vs_SEOB}. Here we consider a more massive BBH with $M=100\,M_\odot$. The mass ratio is $q=1/4$ and the spins are $(\chi_{1z}, \chi_{1p})=(-0.5, 0.4)$ and $(\chi_{2z}, \chi_{2p})=(-0.3, 0.7)$. \teja{We find that even for massive BBHs, \xode\ still consistently agrees better with \seob\ than \xphm\ does}{We find that compared to \xphm, \xode\ still consistently agrees better with \seob\, even for massive BBHs}. On the other hand, as the total mass increases and the inspiral's fractional contribution to the SNR decreases, the overall agreement between both phenomenological models and \seob\ tends to increase. This is again because the precession timescale $t_{\rm pr}$ is longer than the GW decay timescale $t_{\rm gw}$ and the Euler angle $\alpha$ becomes slow varying with respect to frequency (see also the discussion above related to Fig.~\ref{fig:eg_qq_25_chiz_n7_chip_3_vs_NR}). Therefore, for the rest of the paper, we will focus on light BBHs where the need to improve the description of precession is most crucial.

\subsection{Parameter space exploration}
\label{sec:MM_par_space}

\begin{figure*}
  \centering
  \includegraphics[width=1.9\columnwidth]{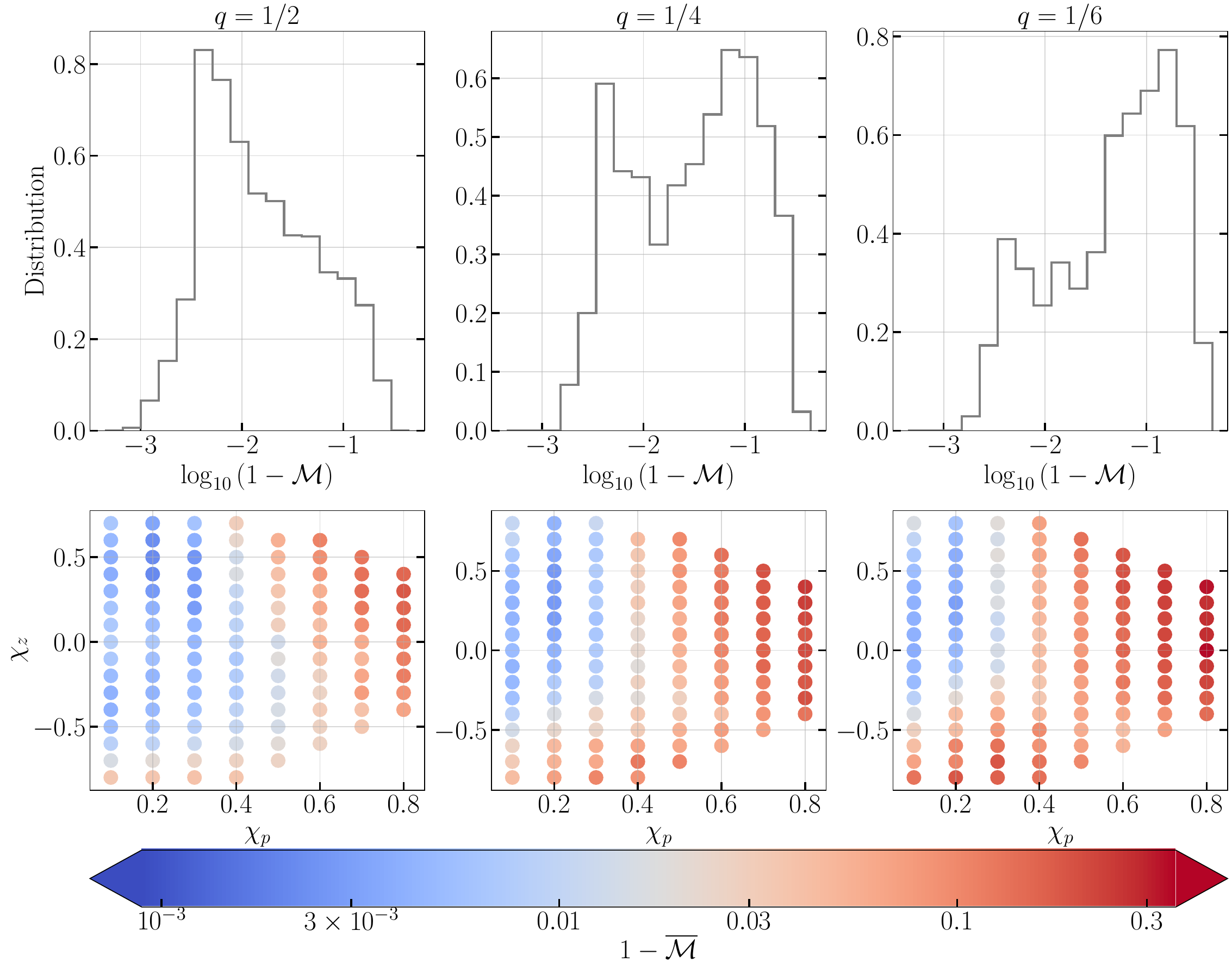}
\caption{A survey of the mismatch between \seob\ and \xphm\ over different mass ratios and spin configurations. We fix $M=15\,M_\odot$ and $\iota = \pi/3$ while randomize other extrinsic parameters. The top row shows the distribution of the mismatch and the bottom row shows the SNR-weighted mismatch at each specific spin configuration. Large mismatches are noted for small $q$, large $\chi_p$, and negative $\chi_z$. 
}
\label{fig:SEOB_vs_XPHM_par_space}
\end{figure*}

\begin{figure*}
  \centering
  \includegraphics[width=1.9\columnwidth]{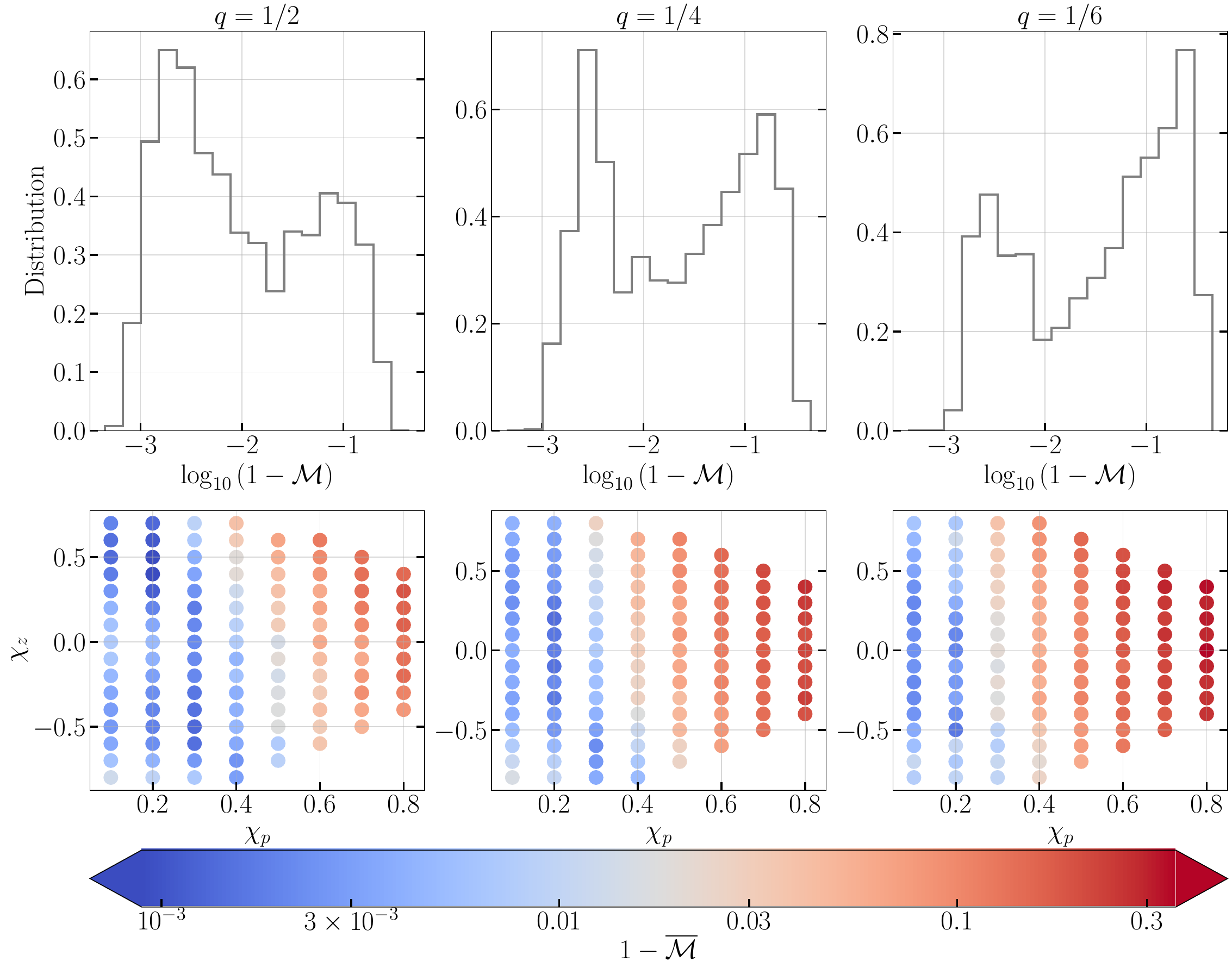}
\caption{A survey of the mismatch between \seob\ and \xode. Here the recalibration of coprecessing mode is turned off, so \xode\ only modifies the Euler angles used by \xphm. This improves the agreement with \seob\ for negative $\chi_z$ and for small $\chi_p\lesssim 0.4$. For larger values of $\chi_p$, the mismatch is still significant and can often exceed 10\%. 
}
\label{fig:SEOB_vs_XODE_par_space}
\end{figure*}

\begin{figure*}
  \centering
  \includegraphics[width=1.9\columnwidth]{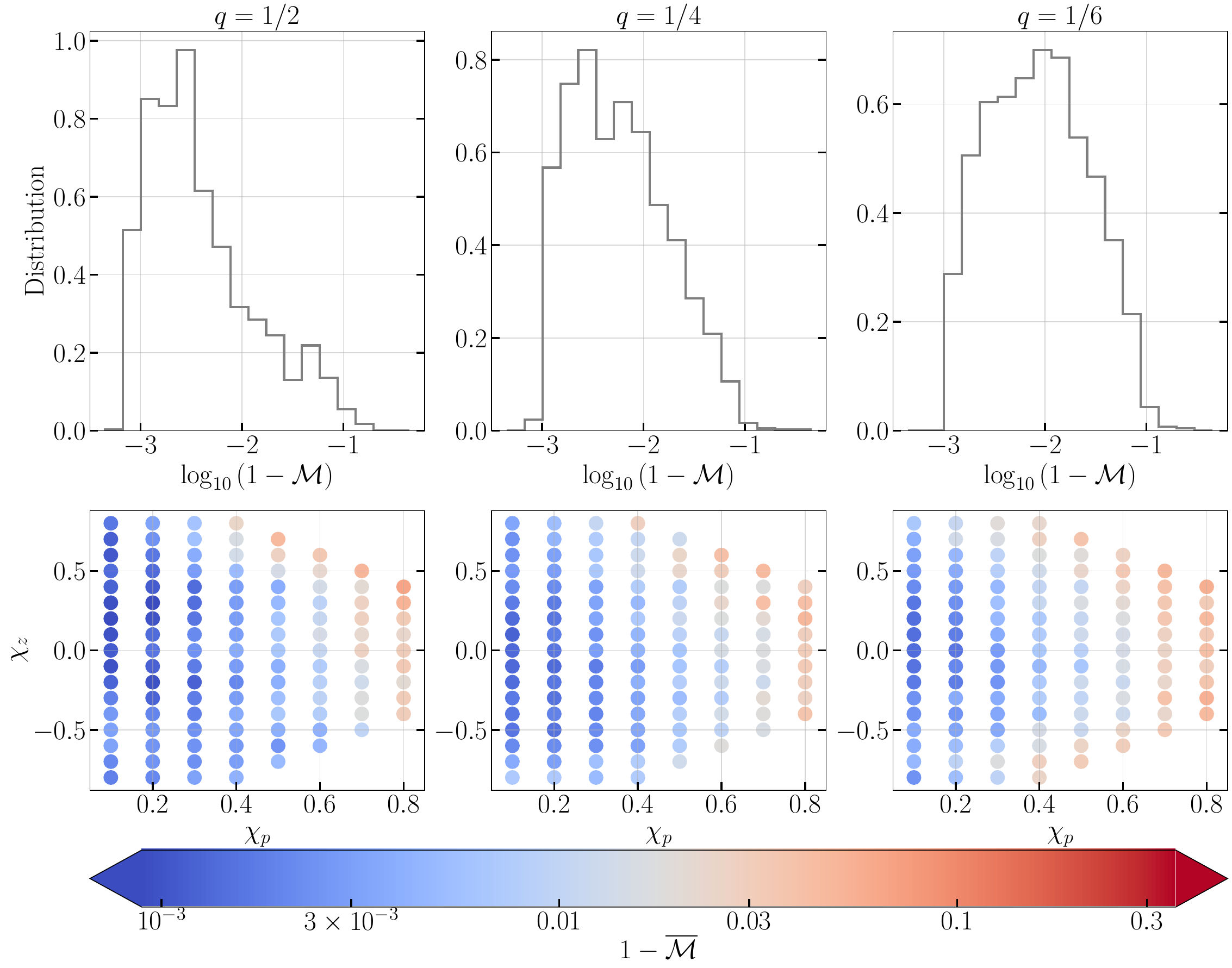}
\caption{Similar to Fig.~\ref{fig:SEOB_vs_XODE_par_space} but now include also the recalibration of coprecessing modes, which is the default behavior of \xode. In this case, we find good agreement between the two approximants for most parts of the parameter space. 
}
\label{fig:SEOB_vs_XODE_cp_cal_par_space}
\end{figure*}

Having examined a few representative examples in Sec.~\ref{sec:MM_eg}, we now systematically explore the consistency between the phenomenological models and \seob\ for three representative mass ratios, $q=1/2$, $1/4$, and $1/6$. In all cases, we fix the total mass $M=15\,M_\odot$. Here we focus on systems with relatively small total masses to emphasize the inspiral part where \seob\ should give an accurate representation of the true waveform (but also see Fig.~\ref{fig:eg_qq_25_chiz_n5_chip_4_vs_SEOB}). 
For each value of $q$, we scan through $\chi_{1z}=\chi_{2z}=(-0.8, ..., 0.8)$ and $\chi_{1p}=\chi_{2p}=(0.1, ..., 0.8)$ while restricting $\chi_1=\chi_2<0.9$. We fix the inclination to be $\iota = \pi/3$ for all of the systems. For each set of intrinsic parameters, we consider 20 different realizations that randomize over $\kappa$, $\phi_{\rm ref}$, as well as the azimuthal orientation of $(\chi_{1p}, \chi_{2p})$. Note that the two spins are not aligned at $f_{\rm ref}$ (=14.5\,Hz) in general as we sample the azimuthal angles of $(\chi_{1p}, \chi_{2p})$ independently. We assume the design sensitivity of  aLIGO~\cite{LSC:15det, Barsotti:18}, and compute matches starting from $f_{\rm low} = 15\,{\rm Hz}$. 

To set a benchmark, we first compare mismatches between \xphm\ and \seob\ in Fig.~\ref{fig:SEOB_vs_XPHM_par_space}. The three columns correspond to the three different mass ratios we consider. The top row shows the SNR-weighted distribution of the log mismatch,\footnote{Note that here we treat each spin configuration as equally likely and ignore the astrophysical likelihood as it is still largely uncertain.} and the bottom row shows the SNR-weighted match, $\overline{\mathcal{M}}$, over the $(\chi_p, \chi_z)$ parameter space, where $\chi_p=\chi_{1p}=\chi_{2p}$ and $\chi_z=\chi_{1z}=\chi_{2z}$ in our study. The definition of $\overline{\mathcal{M}}$ follows Ref.~\cite{Harry:16, Pratten:21}, 
\begin{equation}
    \overline{\mathcal{M}} = \left( \frac{\sum_i \rho_i^3 \mathcal{M}_i^3 }{\sum_i \rho_i^3 }\right)^{1/3}, 
\end{equation}
where $\rho_i$ is the SNR computed using \seob\ for the $i$'th realization. In the plot, the color coding is based on $1-\overline{\mathcal{M}}$. 

From Fig.~\ref{fig:SEOB_vs_XPHM_par_space}, we note that the mismatch between \xphm\ and \seob\ increases as $\chi_p$ increases and as $q$ decreases. Moreover, the mismatch degrades quickly as $\chi_z$ becomes more negative. Whereas the mismatch remains below 3\% until $\chi_p\gtrsim 0.5$ for a BBH with a comparable mass ratio of $q=1/2$, for a BBH with a more asymmetric mass ratio of $q=1/6$, even $\chi_p=0.1$ can lead to a significant mismatch if the spin is also anti-aligned.
Such a system can experience transitional precession~\cite{Apostolatos:94} because the decaying orbital angular momentum $L$ can nearly cancel with the spin angular momentum $S\simeq S_1$, making the total angular momentum nearly zero ($J\simeq 0$) during the inspiral.
Indeed, we find both the MSA and NNLO constructions of the Euler angles quickly become inaccurate as $\chi_z$ becomes more negative (see, e.g., Fig~\ref{fig:MSA_vs_NLO_ODE}). 
Overall, if $q\lesssim 1/4$, a large fraction of systems have mismatches exceeding 10\%. This implies that the systematic error of precessing waveforms is significant and should be properly taken into consideration when we analyze both individual events and the population of BBHs. We will get back to this point again in Sec.~\ref{sec:conclusion_discussions}. 

If we modify the evolution of Euler angles by the numerical solution to the N4LO precession equations but still twist up the same coprecessing modes as used by \xphm, the resultant mismatch against \seob\ is shown in Fig.~\ref{fig:SEOB_vs_XODE_par_space}. Here we notice an improvement over \xphm\ for small to moderate $\chi_p \lesssim 0.4$, and especially for BBHs that will experience transitional precession with large anti-aligned spins. However, we still see large discrepancies between approximants for large $\chi_p$ due to the phase difference of the coprecessing modes (see Figs.~\ref{fig:XODE_vs_SEOB_debug} and \ref{fig:eg_qq_25_chiz_p2_chip_6_vs_SEOB}). The large fraction of BBHs with mismatch greater than $10\%$ means that fixing the Euler angles alone is insufficient for improving the accuracy of phenomenological waveforms.  

Lastly, we present in Fig.~\ref{fig:SEOB_vs_XODE_cp_cal_par_space} the mismatch between \seob\ and \xode\ under the default usage. In other words, we include both the numerical Euler angles and the recalibration of coprecessing modes). In this case, the majority of the BBHs have mismatches between $0.1\%-1\%$ even for very asymmetric systems with $q=1/6$, and rarely do we see systems with mismatches exceeding 10\% [which happens only for systems with large ($\gtrsim 0.8$) and misaligned spins]. We thus conclude that \xode\ serves as an accurate frequency-domain representation of \seob.

\section{Conclusion and discussions}
\label{sec:conclusion_discussions}


In this work, we introduced \xode, a frequency-domain waveform approximant for generically precessing BBHs. We build upon the \xphm\ approximant \cite{Pratten:21} and include two new ingredients. First, we discussed the limitation of the MSA construction for the Euler angles in Sec.~\ref{sec:limitations_p_avg},  which motivated us to instead obtain the Euler angles by numerically integrating the N4LO precession equations. We sped up the numerical integration using an acceleration technique described in Sec.~\ref{sec:fast_prec}, which may also be useful for other approximants that rely on the numerical solutions to the precession equations~\cite{Colleoni:23, Gamba:22, Estelles:22}.
Second, we recalibrated the phase of coprecessing modes to better match those of \seob\ (Sec.~\ref{sec:cp_modes}). The implementation of our waveform is highly efficient, especially for systems with long durations (i.e., high frequency resolutions; Fig.~\ref{fig:evaluation_time}). More importantly, it agrees much better with \seob\ than \xphm\ does, and therefore may potentially mitigate systematic errors in the waveforms at least in the inspiral part where \seob\ is expected to have high accuracy. The Python source code of our waveform is publicly available at \url{https://github.com/hangyu45/IMRPhenomXODE}.  

In our investigations, we noted significant discrepancies between \xphm\ and \seob\ when examining their matches over parameter space covering different mass ratios and spin configurations (Fig.~\ref{fig:SEOB_vs_XPHM_par_space}). If the binary is strongly precessing and has an asymmetric mass ratio, the mismatch can be as large as 10\% or more. The disagreement is especially significant for BBHs with anti-aligned spins.  
This is consistent with the finding that the PE posteriors obtained using these two approximants disagree for some GW events (e.g., GW190412~\cite{LVC:20GW190412}, GW190521~\cite{LVC:20GW190521properties, Estelles:22b}; see also Sec. V E of Ref.~\cite{LVK:21gwtc3}).

The large discrepancies also mean that we should be cautious about systematic errors in waveform models when drawing conclusions about the population properties of BBHs. 
When estimating the selection effects, it is crucial for us to account for not only the SNR but also the waveform accuracy. For example, it has been previously suggested that there are more aligned BBHs than anti-aligned ones~\cite{Roulet:21}. However, this is based on analyzing posteriors obtained from \xphm\ and assuming that the waveform model's accuracy is homogenous across the parameter space. Based on our Fig.~\ref{fig:SEOB_vs_XPHM_par_space}, this may not be the case as \xphm\ shows larger mismatches with \seob\ for anti-aligned systems as compared to aligned ones. Another example is that current population studies suggest that the BBH mergers detected by the LVK favor more symmetric mass ratios, $q\simeq 1$, as compared to asymmetric ones~\cite{LVK:21pop}. Our results also show that different waveform approximants are more discrepant for systems with more asymmetric values of the mass ratio $q$. Hence, it is worth revisiting the population studies taking into account the systematic uncertainties in the waveforms. It would be especially interesting to reexamine the spin distribution of BBHs with \xode. If \seob\ is indeed a better description of the true waveform than \xphm\ (see, e.g., Ref.~\cite{RamosBuades:23}), then Fig.~\ref{fig:SEOB_vs_XODE_cp_cal_par_space} suggests that \xode\ will have a much smaller systematic error compared to \xphm\ (cf. Fig.~\ref{fig:SEOB_vs_XPHM_par_space}). 

Meanwhile, the high efficiency of \xode\ is another factor that makes it suitable for analyzing a large number of systems as required by population studies. We also highlight the fact that \xode\ appears to be more efficient than \xphm\ for evaluating waveforms on large frequency grids with $\gtrsim 10^4$ bins (Fig.~\ref{fig:evaluation_time}). As we expect the sensitivity of ground-based GW detectors to improve steadily at low frequencies where the GW decay timescale is long, the signal below 20\,Hz will become increasingly important, especially for constraining the precession dynamics as most precession cycles happen at low frequencies (Sec.~\ref{sec:limitations_p_avg}). This is especially the case when the next generation of GW detectors (including the Cosmic Explorer~\cite{Evans:17, Evans:21} and the Einstein Telescope~\cite{Hild:10, Sathyaprakash:12}) become available. Therefore, it is crucial for a waveform to be generated efficiently for a BBH with a long time-domain duration and a high frequency-domain resolution (see also, e.g., Ref.~\cite{Thomas:22} for producing computationally efficient waveforms with neural networks). Even in cases where relative binning~\cite{Zackay:18, Cornish:21} is used, \xode\ still has comparable computational efficiency to \xphm\ with a typical evaluation time of 20\,ms per waveform. 

Besides helping the construction of waveforms for precessing BBHs, the acceleration technique in Sec.~\ref{sec:fast_prec} can also be used to evolve a BBH backward in time so that we can obtain the spin configuration at the formation which is more directly related to the astrophysical formation channel ~\cite{JohnsonMcDaniel:22, Gerosa:23}. 
For this purpose, we should improve our analytical baseline estimation so that the residual to be tracked numerically is reduced. Our current choice of Eqs.~(\ref{eq:transformation}) and (\ref{eq:al_0}) is motivated by the simple precession analysis assuming only one BH is spinning~\cite{Apostolatos:94}. It can reduce the number of steps in the ODE integration by about a factor of 2 compared to the direct integration of the original equations, which is sufficient for our purpose to make the computation of the Euler angles subdominant in the waveform generation. However, we ignore spin-spin interactions that can happen at timescales shorter than the GW decay timescale and dominate the numerical computation at low frequencies. Nevertheless, spin-spin interaction can be accounted for if we instead use the more sophisticated MSA construction~\cite{Chatziioannou:17} as a refined baseline estimation. Another ingredient that should be incorporated is the eccentricity of the binary and some recent developments can be found in, e.g., Refs.~\cite{Phukon:19, Yu:20a, Klein:21}.


There are also other directions for us to improve \xode. For example, the current version of \xode\ focuses on improving mainly the accuracy in the inspiral part. We calibrated the coprecessing modes to \seob\ in order to capture effects such as the time-dependent evolution of $(\chi_{1z}, \chi_{2z})$ and potentially a more self-consistent estimation of the final spin. 
During the preparation of this manuscript, an update to \seob, \texttt{SEOBNRv5PHM}~\cite{RamosBuades:23}, has become available. We would thus like to update the recalibration of the coprecessing modes accordingly. 
However, both \seob~\cite{Ossokine:20} and its v5 update~\cite{RamosBuades:23} are not particularly calibrated to precessing NR simulations but only validated against them. 
We would thus like to further integrate \xode\ together with, e.g., \texttt{IMRPhenomPNR}~\cite{Hamilton:21} and its new extension \texttt{IMRPhenomXO4a} \cite{London:23, Thompson:23, Ghosh:23b}. These recently developed approximants are calibrated to NR to capture precession effects in both Euler angles and coprecessing modes in the merger-ringdown phase (see also Ref.~\cite{Hamilton:23}).

Yet another avenue for improvement is to extend \xode\ to incorporate matter effects so that it can be used to capture neutron-star-black-hole binaries and binary neutron stars. If $M_2$ is a spinning neutron star, the spin-induced mass quadrupole will lead to extra precession~\cite{Barker:75, Poisson:98, LaHaye:22} while the tidal effects modify the orbital frequency evolution rate~\cite{Flanagan:08, Hinderer:16, Steinhoff:21, Yu:23a}. 
We anticipate the acceleration technique introduced in Sec.~\ref{sec:fast_prec} will still work when matter effects are present, as long as matter effects can be treated as small perturbations to the main BBH dynamics. 
Besides the smooth components, a spinning NS can also have a rich spectrum of inertial modes that can be resonantly excited during the inspiral~\cite{Flanagan:07, Xu:17, Poisson:20, Ma:21, Gupta:21}, leading to additional features in the waveform that can be utilized to improve our constraints on the system parameters.

\begin{acknowledgments}
We thank Maria Haney, Marta Colleoni, Katerina Chatziioannou, and other LVK colleagues for their useful comments during the preparation of this manuscript. 
HY's work at KITP is supported by the National Science Foundation (NSF PHY-1748958) and by the Simons Foundation (216179, LB). HY is also supported by NSF PHY-2308415. TV acknowledges support from NSF Grants PHY-2012086 and PHY-2309360, the Alfred P. Sloan Foundation through grant number FG-2023-20470, and the Hellman Family Faculty Fellowship.  MZ is supported by NSF 2209991 and NSF-BSF 2207583. Use was made of computational facilities purchased with funds from the National Science Foundation (CNS-1725797) and administered by the Center for Scientific Computing (CSC). The CSC is supported by the California NanoSystems Institute and the Materials Research Science and Engineering Center (MRSEC; NSF DMR 1720256) at UC Santa Barbara.

\end{acknowledgments}

\appendix

\section{Coefficients of \texorpdfstring{$\Lambda$}{Λ}}
\label{appx:Lambda}
We present here in Tables~\ref{tab:coeff_psi_22}-\ref{tab:coeff_psi_33} numerical values of $\Lambda$ in Eq.~(\ref{eq:phs_2_phenom}). Scientific notation will be used in all the tables. 

\begin{table}
\caption{Numerical values for the coefficient matrix $\Lambda$ for the $(\ell, |m'|)=(2, 2)$ coprecessing mode. 
\label{tab:coeff_psi_22}}
\begin{ruledtabular}
\begin{tabular}{c|ccc}
            & $\lambda_0$ & $\lambda_1$ & $\lambda_2$\\
\hline
1                       & -3.56324E-02      &   -3.91800E-01    & -9.75686E-01  \\
$\eta$                  & 6.88385E-01       &   9.66232E+00     & 4.18895E+01   \\
$\eta^2$                & -1.35385E+00      &   -1.83658E+01    & -7.72334E+01  \\
$\chieff$               & -2.54389E-01      &   -3.80333E+00    & -1.83726E+01  \\
$\eta\chieff$           & 2.09265E+00       &   3.03360E+01     & 1.38345E+02   \\
$\eta^2\chieff$         & -4.63154E+00      &   -6.63854E+01    & -2.91346E+02  \\
$\chi_p$                & -1.13199E-01      &   -2.37404E+00    & -1.62808E+01  \\
$\eta \chi_p$           & 2.02797E+00       &   3.74024E+01     & 2.26834E+02   \\
$\eta^2 \chi_p$         & -7.76537E+00      &   -1.37812E+02    & -7.92164E+02  \\
$\chieff \chi_p$        & -1.22152E-01      &   -1.60476E+00    & -4.48829E+00  \\
$\eta \chieff \chi_p$   & 5.91457E-02       &   -4.31768E-01    & -1.67820E+01  \\
$\chi_p^2$              & -2.83718E+00      &   -5.07668E+01    & -2.87187E+02  \\
$\eta \chi_p^2$         & 1.88757E+01       &   3.36199E+02     & 1.90689E+03   \\
$\eta^2 \chi_p^2$       & -3.11665E+01      &   -5.53809E+02    & -3.16054E+03  \\
\end{tabular}
\end{ruledtabular}
\end{table}

\begin{table}
\caption{$\Lambda$ for the $(\ell, |m'|)=(2, 1)$ coprecessing mode. 
\label{tab:coeff_psi_21}}
\begin{ruledtabular}
\begin{tabular}{c|ccc}
            & $\lambda_0$ & $\lambda_1$ & $\lambda_2$\\
\hline
1                       & 1.45776E-01   & 2.18974E+00   & 1.04487E+01   \\
$\eta$                  & -1.40633E+00  & -2.05837E+01  & -9.54318E+01  \\
$\eta^2$                & 3.46278E+00   &  5.03418E+01  & 2.30135E+02   \\
$\chieff$               & 1.29801E-01   & 1.78996E+00   & 7.68368E+00   \\
$\eta\chieff$           & -1.56514E+00  & -2.17761E+01  & -9.42716E+01  \\
$\eta^2\chieff$         & 3.65947E+00   & 5.02751E+01   & 2.13065E+02   \\
$\chi_p$                & -2.36797E-01  & -2.92172E+00  & -1.08634E+01  \\
$\eta \chi_p$           & 3.16000E+00   & 4.03028E+01   & 1.54409E+02   \\
$\eta^2 \chi_p$         & -9.47821E+00  & -1.23387E+02  & -4.81714E+02  \\
$\chieff \chi_p$        & -4.50281E-01  & -6.89852E+00  & -3.36019E+01  \\
$\eta \chieff \chi_p$   & 2.18979E+00   & 3.26216E+01   & 1.55407E+02   \\
$\chi_p^2$              & 5.20529E-02   & -2.57050E+00  & -2.68307E+01  \\
$\eta \chi_p^2$         & -2.11532E+00  & -4.78183E+00  & 1.03294E+02   \\
$\eta^2 \chi_p^2$       & 8.77288E+00   & 7.28983E+01   & 5.46643E+01   \\
\end{tabular}
\end{ruledtabular}
\end{table}

\begin{table}
\caption{$\Lambda$ for the $(\ell, |m'|)=(3, 3)$ coprecessing mode. 
\label{tab:coeff_psi_33}}
\begin{ruledtabular}
\begin{tabular}{c|ccc}
            & $\lambda_0$ & $\lambda_1$ & $\lambda_2$\\
\hline
1                       & -5.16818E-02  & -5.93521E-01  & -1.51008E+00  \\
$\eta$                  & 1.07597E+00   & 1.44729E+01   &  5.64762E+01  \\
$\eta^2$                & -3.13016E+00  & -4.27177E+01  & -1.74271E+02  \\
$\chieff$               & 1.45058E-01   & 2.61181E+00   & 1.41871E+01   \\
$\eta\chieff$           & -2.09505E+00  & -3.66877E+01  & -2.01080E+02  \\
$\eta^2\chieff$         & 6.47145E+00   & 1.11455E+02   & 6.11886E+02   \\
$\chi_p$                & -3.80842E-01  & -5.80229E+00  & -3.09561E+01  \\
$\eta \chi_p$           & 4.82895E+00   & 7.31911E+01   & 3.77033E+02   \\
$\eta^2 \chi_p$         & -1.32012E+01  & -2.02043E+02  & -1.03115E+03  \\
$\chieff \chi_p$        & -2.85049E-01  & -3.63195E+00  & -1.14999E+01  \\
$\eta \chieff \chi_p$   & 3.62108E-01   & 2.76110E+00   & -9.34885E+00  \\
$\chi_p^2$              & -3.11476E+00  & -5.27402E+01  & -2.71699E+02  \\
$\eta \chi_p^2$         & 2.02101E+01   & 3.40495E+02   & 1.75994E+03   \\
$\eta^2 \chi_p^2$       & -3.35661E+01  & -5.62078E+02  & -2.91600E+03  \\
\end{tabular}
\end{ruledtabular}
\end{table}


\bibliography{ref}

\end{document}